\documentclass[aps,prf,reprint,onecolumn,superscriptaddress]{revtex4-2}
\usepackage{graphicx}
\usepackage{picture}
\usepackage{amsmath}
\usepackage{hyperref}
\usepackage{graphicx}
\usepackage[]{color}

\newcommand{\Q}{\mathcal{Q}}
\newcommand{\de}{\mathrm{d}}
\usepackage{xcolor}

\graphicspath{{./figures/}}

\begin{document}


\title{Effect of surfactant-laden droplets on \\ turbulent flow topology }


\author{Giovanni Soligo}
\affiliation{Institute of Fluid Mechanics and Heat Transfer, TU-Wien, 1060 Vienna, Austria}
\affiliation{Polytechnic Department, University of Udine, 33100 Udine, Italy}
\author{Alessio Roccon}
\affiliation{Institute of Fluid Mechanics and Heat Transfer, TU-Wien, 1060 Vienna, Austria}
\affiliation{Polytechnic Department, University of Udine, 33100 Udine, Italy}
\author{Alfredo Soldati}
\affiliation{Institute of Fluid Mechanics and Heat Transfer, TU-Wien, 1060 Vienna, Austria}
\affiliation{Polytechnic Department, University of Udine, 33100 Udine, Italy}


\date{\today}

\begin{abstract}
In this work, we investigate flow topology modifications produced by a swarm of large surfactant-laden droplets released in a turbulent channel flow. Droplets have same density and viscosity of the carrier fluid, so that only surface tension effects are considered. We run one single-phase flow simulation at $Re_\tau=\rho u_\tau h / \mu = 300$, and ten droplet-laden simulations at the same $Re_\tau$ with a constant volume fraction equal to $\Phi \simeq5.4\%$. For each simulation, we vary the Weber number ($We$, ratio between inertial and surface tension forces) and the elasticity number ($\beta_s$, parameter that quantifies the surface tension reduction). We use direct numerical simulations of turbulence coupled with a phase field method to investigate the role of capillary forces (normal to the interface) and Marangoni forces (tangential to the interface)  on turbulence (inside and outside the droplets). As expected, due to the low volume fraction of droplets, we observe minor modifications in the macroscopic flow statistics. However, we observe major modifications of the vorticity at the interface and important changes in the local flow topology. We highlight the role of Marangoni forces in promoting an elongational type of flow in the dispersed phase and at the interface. We provide detailed statistical quantification of these local changes as a function of the Weber number and elasticity number, which may be useful for simplified models.
\end{abstract}



\maketitle

\section{Introduction}

The evaluation of mass, momentum and heat transfers across a deformable interface is a crucially important design parameter in many industrial applications \cite{kralova2009,rosen2012surfactants}.
Usual engineering practice is, however, based on strong simplifying assumptions of the flow field in the proximity of the interface \cite{akita1974bubble,kelly1982interfacial,delhaye1994interfacial}. 
The development of correlations able to accurately predict these transfer rates is extremely difficult due to the limited amount of experimental and numerical data available.
Indeed, obtaining accurate information on physical quantities of interest near a deformable interface is challenging:  experimental measurements on multiphase flows require the development of specific methods \cite{lindken1999velocity,reinecke1998tomographic} or a combination of them \cite{lindken2002novel}, while accurate simulations of multiphase flows require numerical methods able to capture the topology of the dispersed phase and its changes, e.g. coalescence and breakage events \cite{elghobashi2019direct,cmmf2009,TRYGGVASON2011}.

The quest for this type of information becomes even more problematic when real-life environmental and industrial applications are considered and, thus, the presence of impurities and/or surface-active agents (surfactants) has to be accounted for \cite{rosen2012surfactants,pereira2018reduced}.
These agents, which collect at the interface between the two phases, locally reduce surface tension and might generate surface tension gradients, strongly modifying the system evolution \cite{LuMT_2017,Soligo2019_jfm,takagi2008effects,takagi2011surfactant}.
This further complexity requires the adoption of more sophisticated experimental \cite{bull1999surfactant,dussaud2005spreading} and numerical techniques able to describe the surfactant dynamics \cite{adami2010conservative,Soligo2019_jcp,xu2006,MuradogluT_2008}.
In this work, in an attempt to lay useful guidelines for the development of physics-aware empirical correlations, we aim at characterizing the flow modifications produced by a swarm of surfactant-laden droplets released in a turbulent flow.

The study of the flow modifications produced by a dispersed phase drew the attention of many researchers in recent years, since the presence of deformable droplets or bubbles (and of the accompanying surface tension forces) can generate strong modifications in the carrier flow \cite{LU2005,Scarbolo2014,verschoof2016bubble,risso2018}.
In this context, the role of bubble/droplet interface is crucial since it introduces an elasticity element in the system and can modulate the transfer of momentum between the two phases and thus modifies the overall flow dynamics.
Previous numerical works studied different aspects of the problem: Tryggvason and co-authors  \cite{LU2005,lu2006,Lu2007,Lu2008,LuMT_2017,Lu2018,Lu2019} investigated the macroscopic flow modifications produced by a swarm of bubbles; Ferrante and co-authors \cite{Dodd2016,freund2019} studied the effects of droplets properties on turbulence decay; the group of Brandt and co-authors \cite{Devita2019,RostiBM_2018,RostiDB_2019,RostiGJDB_2019} investigated the effects of coalescence and volume fraction on system rheology; \citet{mukherjee2019} and \citet{shao2018direct} studied the influence of volume fraction and Weber number on homogeneous isotropic turbulence; finally, \citet{Scarbolo2016} and \citet{spandan2018physical} investigated the effects of droplets deformability on drag.

In this work, we aim at characterizing the flow modifications introduced by a swarm of surfactant-laden droplets released in a turbulent channel flow.
Due to the relatively low volume fraction of the dispersed phase ($\Phi \simeq 5.4\%$), important macroscopic flow modifications are not expected and indeed we find an extremely limited increase of the flow-rate ($\simeq 1\%$) for all cases (with respect to the single-phase case).
However, important local flow modifications due to the presence of the interfaces are expected.
The interface is an elastic boundary that separates the two phases and confines the flow within, and on top of that, the soluble surfactant introduces additional effects locally reducing the surface tension according to its concentration and strength, and possibly generating forces tangential to the interface (Marangoni forces).
To characterize these local modifications, we analyze the vorticity field at the interface and we study the spatial distribution of the flow topology parameter, $\Q$, in the different regions of the domain: carrier phase, dispersed phase and interface. 
The topology parameter locally defines the type of flow (purely rotational, pure shear and purely elongational flow) based on the symmetric and antisymmetric part of the velocity gradient tensor (i.e. the rate-of-deformation and the rate-of-rotation tensors, respectively).

The analyses presented here are based on the simulations database developed in \citet{Soligo2019_jfm}, in which a two-order-parameter phase-field method \cite{Komura1997,Engblom2013,Yun2014,Soligo2019_jcp} is adopted to describe the interfacial phenomena.
The first order parameter, $\phi$, defines the morphology of the interface, while the second order parameter, $\psi$, describes the dynamics of the soluble surfactant.
For all the simulations, the shear Reynolds number (ratio between inertial and viscous forces) is kept fixed at $Re_\tau=300$.
We consider two different values of the Weber number: $We=1.50$ (high surface tension) and $We=3.00$ (low surface tension) and four different surfactant strengths: from $\beta_s=0.50$ (weaker surfactant) up to $\beta_s=4.00$ (stronger surfactant).
The simulation database is completed by two simulations of surfactant-free systems ($We=1.50$ and $We=3.00$) and by a single-phase flow simulation.

The paper is organized as follows: at first, the numerical method and the simulation set-up are introduced in section~\ref{sec: method}.
Then, the effects introduced by a surfactant-laden and deformable interface are presented in section~\ref{sec: results} for different surface tension values and surfactant strengths.
The last section, section~\ref{sec: conclusions}, summarizes the presented results.

\section{Methodology}
\label{sec: method}

The dynamics of a swarm of surfactant-laden droplets in wall-bounded turbulence is described employing direct numerical simulations of the Navier-Stokes equations coupled with a two-order-parameter phase-field method \cite{Komura1997,Engblom2013,Yun2014,Soligo2019_jcp,Soligo2019_jfm}.
The Navier-Stokes equations, which include an additional term to account for the presence of a surfactant-laden interface, define the hydrodynamics of the system.
The two-order-parameter phase-field method, instead, describes the transport of the phase field and of the surfactant concentration.
In the following sections, first the phase-field method in its two-order-parameter formulation is introduced; then, the force coupling with the Navier-Stokes equations is described.

\subsection{Modeling of interface and surfactant concentration dynamics}

The two-order-parameter phase-field formulation uses two scalar fields (order parameters) to define the transport of the phase field and of the surfactant concentration.
The phase field, $\phi$, defines the local concentration of the two phases; it varies between $\phi=-1$ in the carrier phase and $\phi=+1$ in the dispersed phase. Across the interface, the phase field changes smoothly between the two bulk values; the interface is defined as the iso-level $\phi=0$.
The second order parameter, $\psi$, defines the local surfactant concentration in the entire domain \citep{Laradji1992,Komura1997,Engblom2013,Soligo2019_jcp}: here a soluble surfactant is used, thus, in addition to the dynamics of surfactant over the interface, also adsorption and desorption of surfactant to/from the interface are taken into account.
Surfactant molecules are amphiphilic (i.e. molecules composed by a hydrophilic head and a hydrophobic tail), thus they are preferentially found at the interface between different phases.
Indeed, the surfactant concentration has a low uniform value in the bulk of the phases (surfactant bulk concentration, $\psi_b$) and reaches its maximum value at the interface.
In the following all equations are in dimensionless form; the non-dimensional procedure can be found in our previous works \cite{Soligo2019_jcp,Soligo2019_jfm}.
Two separate Cahn-Hilliard-like equations define the dynamics of the phase field and of the surfactant concentration:
\begin{equation}
\frac{\partial \phi}{\partial t}+\mathbf{u} \cdot \nabla \phi=\frac{1}{Pe_\phi}\nabla^2 \mu_\phi + f_p \, ,
\label{eq: phi}
\end{equation}
\begin{equation}
\frac{\partial \psi}{\partial t}+\mathbf{u} \cdot \nabla \psi=\frac{1}{Pe_\psi}\nabla \cdot [\psi(1-\psi) \nabla \mu_\psi] \, ,
\label{eq: psi}
\end{equation}
where $\mathbf{u}=(u,v,w)$ is the velocity field (streamwise, spanwise and wall-normal components), $Pe_\phi$ and $Pe_\psi$ are the phase field and surfactant P\'eclet numbers, $\mu_\phi$ and $\mu_\psi$ their respective chemical potentials and $f_p$ is the penalty flux introduced with the profile-corrected formulation of the phase-field method \cite{li2016phase,zhang2017,Soligo2018}.
The profile-corrected formulation constitutes an improvement to the standard phase-field formulation: it allows to maintain the equilibrium interfacial profile and it overcomes the drawbacks of the method (e.g. mass leakages among the phases and misrepresentation of the interfacial profile \cite{YUE2007,li2016phase}).
The penalty flux is defined as:
\begin{equation}
f_p=\frac{\lambda}{Pe_\phi} \left[ \nabla^2 \phi -\frac{1}{\sqrt{2}Ch} \nabla \cdot  \left( (1-\phi^2) \frac{\nabla\phi}{|\nabla\phi|} \right) \right] \, ,
\end{equation}
where the numerical parameter $\lambda$ can be set via the scaling $\lambda=0.0625/Ch$ \cite{Soligo2018}. 
The Cahn number, $Ch$, is the dimensionless parameter which defines the thickness of the thin interfacial layer. 
The P\'eclet number is a dimensionless quantity, which defines the ratio of diffusive over convective time-scales for the phase field and the surfactant concentration respectively.

The chemical potentials are derived from a two-order-parameter Ginzburg-Landau free energy functional, $\mathcal{F}[\phi, \nabla \phi, \psi]$.
This functional is composed by the sum of five different contributions:
\begin{equation}
\mathcal{F}[\phi, \nabla \phi, \psi]=\int_\Omega ( f_0 +f_{m} + f_\psi + f_1 + f_b) \de \Omega \, ,
\end{equation}
where $\Omega$ is the domain considered. 
The five different contributions are defined as:
\begin{eqnarray}
f_0 &=&\frac{1}{4}(\phi-1)^2(\phi+1)^2 \, , \\[1pt]
f_m&=&\frac{Ch^2}{2}|\nabla\phi|^2 \,  ,\\[1pt]
f_\psi&=&Pi \left[\psi\log\psi +(1-\psi)\log(1-\psi) \right] \, ,\\[1pt]
f_1&=&-\frac{1}{2}\psi(1-\phi^2)^2  \, , \\[1pt]
f_b&=&\frac{1}{2 E_x}\phi^2\psi \, .
\end{eqnarray}
The first two contributions, $f_0$ and $f_m$, are a function of the phase field variable alone and are the same used in the single-order-parameter phase-field formulation \cite{Scarbolo2015,Roccon2017}.
The term $f_0$ defines the tendency of the multiphase system to separate in two pure phases; this phobic behavior is described by a double-well potential, with minima corresponding to the pure phases.
The interfacial term, $f_m$, accounts for the energy stored in the interfacial layer (i.e. the surface tension); it allows for a limited mixing between the two phases at the interface.
The three latter terms define the dynamics of the surfactant:  the entropy term, $f_\psi$, the surfactant adsorption term, $f_1$, and the bulk term, $f_b$.
The entropy reduction obtained when surfactant is uniformly distributed throughout the entire domain is accounted for by the entropy term. The temperature-dependent parameter $Pi$ defines the diffusivity of surfactant: higher values of $Pi$ (higher temperatures) favor a more uniform surfactant distribution in the domain.
This term strictly limits the surfactant concentration between $\psi=0$ (absence of surfactant) and $\psi=1$ (saturation of surfactant).
The surfactant adsorption term defines the amphiphilic behavior of the surfactant: it is maximum at the interface, while it vanishes in the bulk of the phases.
The last contribution, the bulk term, controls the solubility of the surfactant in the bulk of the phases; it vanishes at the interface and it is maximum in the bulk of the phases.
The surfactant solubility parameter, $E_x$, quantifies the solubility of the surfactant: highly soluble surfactant are characterized by high values of $E_x$.

The phase field and surfactant chemical potentials are obtained by taking the variational derivative of the free energy functional with respect to the two order parameters:
\begin{equation}
\mu_\phi=\frac{\delta \mathcal{F}[\phi,\nabla\phi,\psi]}{\delta \phi}=\phi^3-\phi-Ch^2\nabla^2\phi + \underbrace{Ch^2(\psi\nabla^2\phi+\nabla\psi\cdot\nabla\phi)+\frac{1}{E_x}\phi\psi}_{C_{\phi \psi}} \, ,
\label{eq: muphi}
\end{equation} 
\begin{equation}
\mu_\psi=\frac{\delta \mathcal{F}[\phi,\nabla\phi,\psi]}{\delta \psi}=Pi \log\left(\frac{\psi}{1-\psi}\right)-\frac{(1-\phi^2)^2}{2}+\frac{\phi^2}{2 E_x}  \, .
\label{eq: mupsi}
\end{equation}
As discussed in previous literature works \cite{Engblom2013,Yun2014}, the term $C_{\phi \psi}$ in the phase field chemical potential might introduce an unphysical behavior of the thin interfacial layer; on this basis, it is neglected to restore the correct interfacial dynamics \cite{Yun2014}.
Surface tension forces, then, must be calculated using a geometrical approach \cite{popinet2018}, as an alternative to the thermodynamic approach based on the chemical potential gradients \cite{Laradji1992,Engblom2013}.
The use of a geometrical approach brings in also an additional advantage: the effect of surfactant on surface tension can be selected via a suitable surface tension equation of state.
Thus, surface tension forces are calculated here with a geometrical approach, which relies on the phase field for the computation of the local curvature and on a surface tension equation of state accounting for the effect of surfactant.

From the expression of the two chemical potentials, the equilibrium profiles of the order parameters can be obtained: at the equilibrium, the chemical potentials are uniform in all the domain.
%
For a planar interface located at $s=0$, with $s$ being a coordinate normal to the interface, an analytic equilibrium profile can be derived for both the phase field and the surfactant concentration:
\begin{equation}
\phi(s)=\tanh \left( \frac{s}{\sqrt{2}Ch} \right)\, ,
\label{eq: phi_eq}
\end{equation}
\begin{equation}
\psi(\phi)=\frac{\psi_b}{\psi_b+\psi_c(\phi) (1-\psi_b)}\, .
\label{eq: psi_eq}
\end{equation}
The auxiliary variable, $\psi_c$, depends on the phase field only:
\begin{equation}
\psi_c(\phi)=\exp\left[-\frac{1-\phi^2}{2 Pi}\left(1-\phi^2+\frac{1}{E_x}\right) \right]\, .
\label{eq: psic}
\end{equation}
The phase field is uniform in the bulk of the phases, $\phi=-1$ in the carrier phase and $\phi=+1$ in the dispersed phase, and it follows a hyperbolic tangent profile across the interface.
In a similar way, the surfactant concentration is uniform in the bulk of the phases, equal to the surfactant bulk concentration $\psi_b$, and it peaks at the interface, where surfactant molecules preferentially accumulate.
The maximum value of the surfactant concentration depends on the surfactant parameters $Pi$ and $E_x$ and on the surfactant bulk concentration $\psi_b$, as shown in equations~(\ref{eq: psi_eq})-(\ref{eq: psic}).
The equilibrium profiles for the two order parameters are reported in figure~\ref{fig: eos}(\textit{a}): the phase field is reported in red (linear scale on the left) and the surfactant concentration in blue (logarithmic scale on the right).

\subsection{Hydrodynamics}
\label{sec: hydro}

The hydrodynamics of the multiphase system is described using a one-fluid formulation, in which the effects of a surfactant-laden interface are introduced via an additional interfacial term in the Navier-Stokes equations.
In particular, surface tension forces are computed with a geometrical approach \cite{Yun2014}: the Korteweg tensor \cite{KORTEWEG1901} is used to compute the curvature of the interface while an equation of state is used to define the surfactant action on the surface tension.
In this work, to highlight the effects of the surfactant, two phases with matched density ($\rho_1=\rho_2=\rho$) and matched viscosity ($\mu_1=\mu_2=\mu$) are considered.
With these hypotheses, the mass and momentum conservation equations can be written as follows (dimensionless form):
\begin{equation}
\nabla \cdot \mathbf{u}=0 \, ,
\label{eq: cont}
\end{equation}
\begin{equation}
\frac{\partial \mathbf{u}}{\partial t}+\mathbf{u} \cdot \nabla \mathbf{u}=
-\nabla p +\frac{1}{Re_\tau} \nabla^2 \mathbf{u}+ \frac{3}{\sqrt{8}}\frac{Ch}{We} \nabla \cdot[ f_\sigma(\psi) \mathsf{T_c}]  \, ,
\label{eq: ns}
\end{equation}
where $\mathbf{u}=(u,v,w)$ is the velocity field, $\nabla p$ is the pressure gradient (sum of a mean component that drives the flow and of a fluctuating part), $f_\sigma(\psi)$ is the surface tension equation of state \cite{chang1995adsorption,bazhlekov2006numerical} and $\mathsf{T_c}=|\nabla\phi|^2 \mathsf{I}-\nabla\phi\otimes \nabla \phi$ is the Korteweg tensor \cite{KORTEWEG1901}.
The last term in the Navier-Stokes equations, which includes the surface tension equation of state and the Korteweg tensor, accounts for the surface tension forces introduced by the surfactant-laden interface.
This term can be divided into two separate contributions, one normal to the interface (capillary forces) and one tangential to the interface (Marangoni forces).
\begin{equation}
\nabla\cdot[ f_\sigma(\psi) \mathsf{T_c}]= \underbrace{f_\sigma(\psi) \nabla \cdot \mathsf{T_c}}_{\text{Normal (capillary)}} + \underbrace{ \nabla f_\sigma (\psi)  \cdot  \mathsf{T_c} \,}_{\text{Tangential (Marangoni)}}
\label{eq: split}
\end{equation}
When the surface tension is uniform, the tangential contribution vanishes, leaving only the normal contribution (capillary forces).
The shear Reynolds number $Re_\tau=\rho u_\tau h/ \mu$ is defined using the friction velocity $u_\tau=\sqrt{\tau_w/\rho}$ as velocity scale (where $\tau_w$ is the mean wall-shear stress) and the channel half-height, $h$, as length scale.
The Weber number $We=\rho u_\tau^2 h / \sigma_0$ is defined with the surface tension of a clean interface (absence of surfactant), $\sigma_0$, as a reference.

\begin{figure}
\center
\setlength{\unitlength}{0.0025\columnwidth}
\begin{picture}(400,160)
\put(-10,-10){\includegraphics[width=0.6\columnwidth, keepaspectratio]{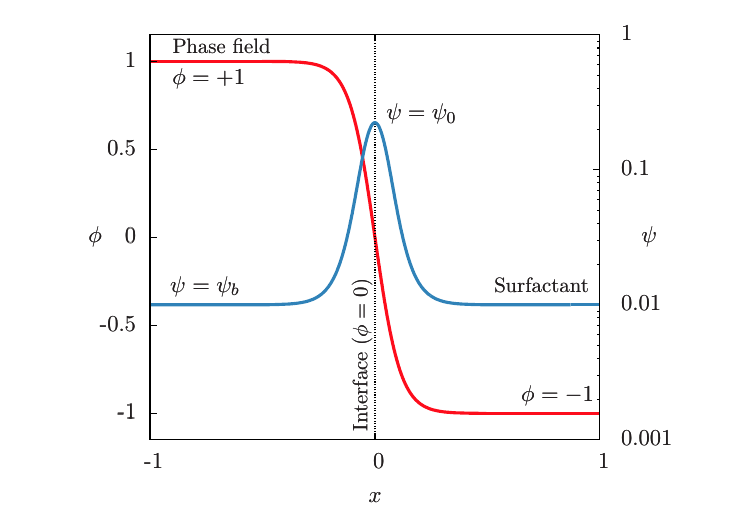}}
\put(195,-10){\includegraphics[width=0.6\columnwidth, keepaspectratio]{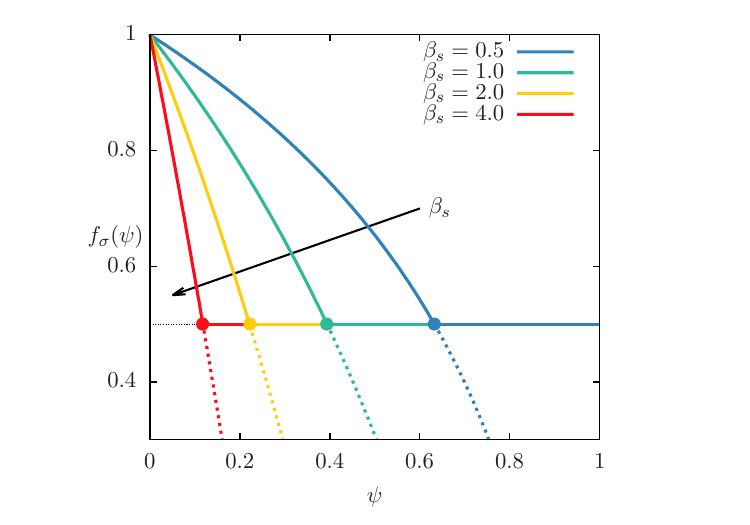}}
\put(5,135){(a)}
\put(205,135){(b)}
\end{picture}
\caption{Panel (a) shows the equilibrium profiles for the phase field (red solid line, linear scale on the left) and for the surfactant concentration (blue solid line, logarithmic scale base 10 on the right).
Panel (b) shows the modified Langmuir equation of state for different elasticity numbers (solid lines).
To remove unphysical surface tension values, the surface tension reduction is limited to $f_\sigma=0.5$ (colored dots).
The behavior predicted by the original Langmuir equation of state is reported with dashed lines below the minimum surface tension, $f_\sigma=0.5$.}
\label{fig: eos}
\end{figure}

The surface tension equation of state adopted here is a modified Langmuir equation of state \cite{Szyszkowski1908}.
Experimental measurements on liquid/liquid \cite{LopezGV_2006,JuJGWZ_2017} and gas/liquid \cite{chang1995adsorption} systems showed that surface tension never decreases below about half its clean value for any surfactant concentration.
This characteristic originates from the saturation of the interface: once the surfactant saturation concentration is reached at the interface, no more surfactant molecules can accumulate there, thus surface tension keeps constant \cite{Porter1991}.
This behavior is not captured in the original formulation, which well follows experimental measurements at low surfactant concentrations, but fails to reproduce the surface tension plateau above the surfactant saturation concentration.
For this reason, we adopt a modified equation of state, in which the minimum surface tension value has been limited according to experimental measurements: here the minimum value is set to $0.5\sigma_0$, which is within the range of values measured in experiments (30\% to 60\% \cite{chang1995adsorption,LopezGV_2006,JuJGWZ_2017} of the clean value, $\sigma_0$).
The modified Langmuir equation of state, thus, combines the accuracy of the original equation of state at surfactant concentrations lower than the saturation value with the surface tension saturation observed in experiments.
The equation reads as:
\begin{equation}
f_\sigma(\psi)=\frac{\sigma(\psi)}{\sigma_0}=\max\left[ \underbrace{1+\beta_s \log\left(1-\psi\right)}_{\text{Langmuir EOS}} , 0.5\right] \, .
\label{eq: sigma}
\end{equation}

A similar equation of state has also been employed in previous works \cite{MuradogluT_2008,MuradogluT_2014}; the authors, however, selected a much lower minimum surface tension value to keep the surface tension value strictly positive.
The equation of state is made dimensionless using the surface tension of a clean interface as a reference: when surfactant is absent ($\psi=0$), the dimensionless surface tension results in $f_\sigma(\psi=0)=\sigma(\psi=0)/\sigma_0=1$.
The elasticity number, $\beta_s$, quantifies the strength of the surfactant: higher values of the elasticity number lead to a stronger reduction in the surface tension for the same surfactant concentration.
The modified Langmuir equation of state for the four elasticity numbers used in this work is reported in figure~\ref{fig: eos}(b).

The surfactant saturation value (filled dots in figure~\ref{fig: eos}b), above which surface tension keeps constant at $f_\sigma(\psi)=0.5$,  is here referred to as shutdown concentration, $\psi_s$, and can be calculated as:
\begin{equation}
\psi_s(\beta_s)=1-e^{-0.5/\beta_s}   \; .
\end{equation}
The shutdown concentration is an important parameter that influences the action of the Marangoni forces: these forces are proportional to the surface tension gradients (equation~\ref{eq: split}), thus they vanish whenever surface tension is uniform (i.e. when $\psi >\psi_s$ or when a clean system is considered).

Finally, it is worth to mention that the accumulation of the surfactant molecules at the interface introduces also an additional effect: the interface viscosity \cite{Langevin2014,elfring2016}.
This effect is however negligible for soluble surfactants \cite{Langevin2014}, and thus it is not here considered.

\subsection{Numerical method}

The governing equations~(\ref{eq: phi}), (\ref{eq: psi}), (\ref{eq: cont}) and (\ref{eq: ns}) are solved in a closed channel geometry using a pseudo-spectral method \cite{HussainiZ1987,CanutoHQZ1988,Peyret2002}. 
Fourier series are used to discretize the variables in the two homogenous directions (streamwise, $x$, and spanwise, $y$), while Chebyshev polynomials are used along the wall-normal direction, $z$.
All the five unknowns (velocity, phase field and surfactant concentration) are Eulerian fields defined on the same Cartesian grid (uniform spacing in the homogeneous directions, Chebyshev-Gauss-Lobatto points in the wall-normal direction).

The time advancement of the equations is performed using an implicit-explicit integration scheme.
The linear diffusive term of the equations is discretized with an implicit scheme: Crank-Nicolson for the Navier-Stokes equations and implicit Euler for the phase field and surfactant transport equations. 
This latter scheme damps eventual unphysical high-frequency oscillations that could originate from the steep gradients present in the two Cahn-Hilliard-like equations \cite{Badalassi2003,YUE2004}. 
By opposite, the non-linear part is integrated explicitly over time using a two-step Adams-Bashforth scheme.

A velocity-vorticity formulation is adopted to avoid the costly computation of the pressure field \cite{KIM1987,Speziale1987}: the Navier-Stokes equations are recast in a $4^{th}$ order equation for the wall-normal component of the velocity (twice the curl on the Navier-Stokes equations) and a $2^{nd}$ order equation for the wall-normal component of the vorticity $\omega_z$ (curl on the Navier-Stokes equations).
This set of equations in four unknowns (velocity and wall-normal vorticity) is completed by the continuity equation (mass conservation) and by the definition of wall-normal vorticity.
The phase field transport equation ($4^{th}$ order) is split in two second order equations \cite{YUE2004} while the surfactant transport equation ($2^{nd}$ order) is directly solved in its original formulation \cite{Soligo2018,Soligo2019_jcp,Soligo2019_jfm}.

\subsubsection{Boundary conditions}
For the Navier-Stokes equations, no-slip boundary conditions are applied on the velocity at both solid walls, $z/h=\pm1$:
\begin{equation}
\mathbf{u}(z/h=\pm1)=\mathbf{0} \; , \qquad \frac{\partial w}{\partial z}(z/h \pm 1)=0 \, ;
\end{equation}
the no-flux condition comes from the continuity equation at the wall.
From the no-slip conditions on the velocity, the boundary conditions for the wall-normal component of the vorticity are obtained:
\begin{equation}
\omega_z(z/h=\pm1)=\left.\frac{\partial v}{\partial x}\right|_{z/h=\pm1} -\left.\frac{\partial u}{\partial y}\right|_{z/h=\pm1} =0
\end{equation}
No-flux boundary conditions are imposed on the phase field, surfactant concentration and respective chemical potentials at the two solid walls.
This is formally equivalent to the following boundary conditions: 
\begin{equation}
\left.\frac{\partial \phi}{\partial z}\right|_{z/h=\pm1}=0 \; ,  \quad
\left.\frac{\partial^3 \phi}{\partial z^3} \right|_{z/h=\pm1}=0 \; ,  \quad
\left.\frac{\partial \psi}{\partial z} \right|_{z/h=\pm1}=0  \, .
\end{equation}
Lastly, periodic boundary conditions are applied to all variables along the homogeneous directions (Fourier discretization).
The set of boundary conditions employed strictly enforces the conservation of the total mass of the two phases and the surfactant over time:
\begin{equation}
\frac{\partial}{\partial t} \int_\Omega \phi \de \Omega =0 \; , \quad 
\frac{\partial}{\partial t} \int_\Omega \psi \de \Omega =0 \, .
\end{equation}
Despite the total mass of the two phases is conserved, mass conservation of each phase is not guaranteed \cite{YUE2007,Soligo2018}: limited mass leakages among the phases may occur.
This issue is rooted in the phase-field method and can be strongly mitigated with the adoption of corrected formulations \cite{li2016phase,Soligo2018,zhang2017}; in particular, here the profile-corrected formulation is used \cite{li2016phase,Soligo2018}.
In the cases presented here, mass leakages are limited to at most 8\% of the dispersed phase and occur only in the initial transient phase; at steady-state, the mass of each phase keeps constant.
The mass leakage in the initial part of the simulations is linked with the initial condition chosen and will be better addressed in the following section.

\subsection{Simulation setup}
\label{sec: setup}

\begin{figure}
\center
\setlength{\unitlength}{0.0025\columnwidth}
\begin{picture}(400,180)
\put(20,0){\includegraphics[width=0.9\columnwidth, keepaspectratio]{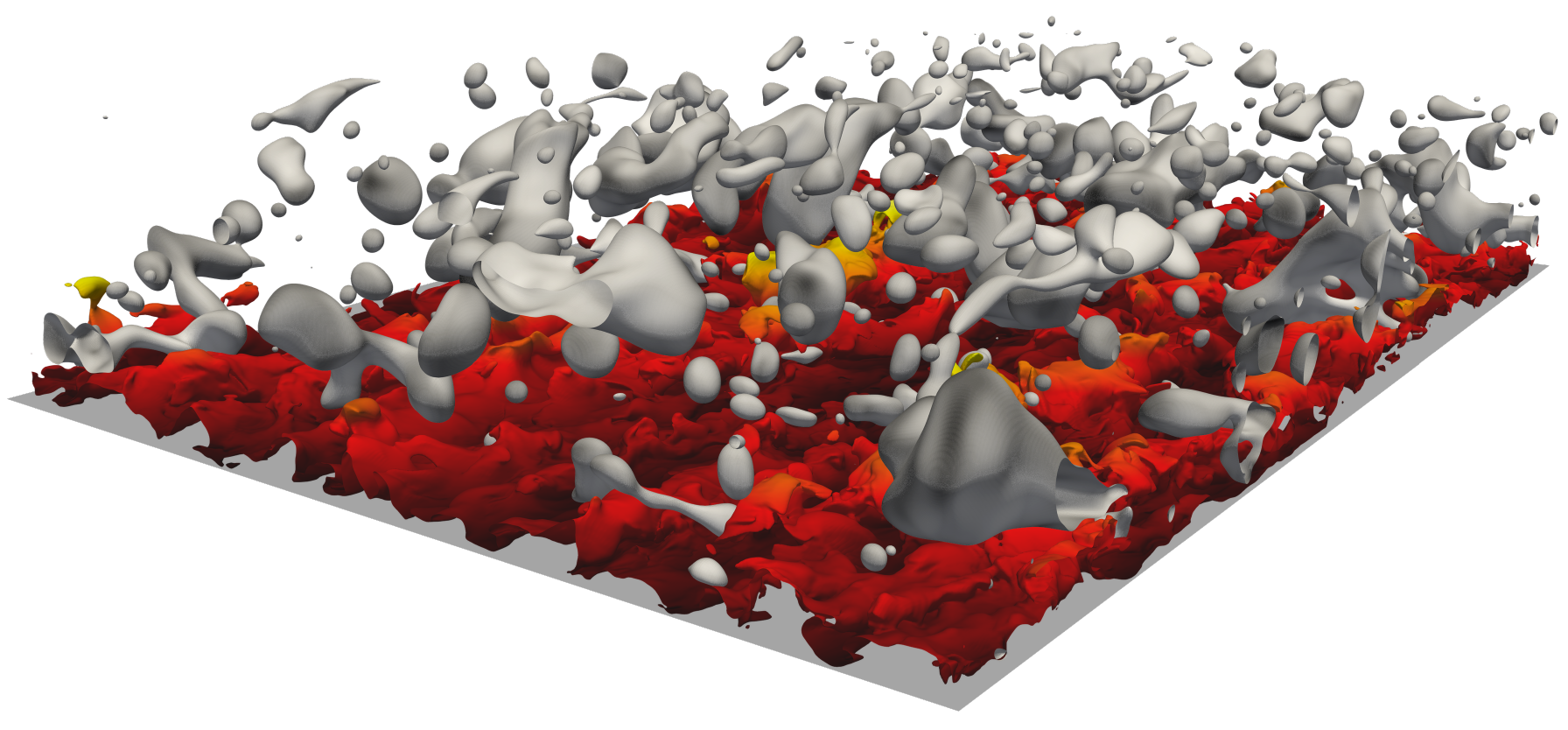}}
\end{picture}
\caption{Instantaneous representation of the droplets at $t^+=3750$ for the case $We=3.00$ and $\beta_s=4.00$.
The interface of the droplets (iso-contour $\phi=0$) is colored by the local surfactant concentration (white-low, black-high). 
The flow field is visualized via the streamwise velocity iso-surface (corresponding to $u = 15$) and is colored by the distance from the bottom wall (black-near wall, yellow-channel centre).
For the sake of clarity, the top wall has been removed and the flow field is shown only in the bottom half of the channel.}
\label{fig: pictorial}
\end{figure}

The same simulation setup adopted in \citet{Soligo2019_jfm} is used in this work, table~\ref{overview}.
A closed channel configuration with dimensions  $L_x \times L_y \times L_z= 4\pi h   \times 2\pi h \times 2h$ ($L_x^+ \times L_y^+ \times L_z^+=3770 \times 1885 \times 600 $ wall units) has been selected, see figure~\ref{fig: pictorial}.
The computational domain is discretized with $N_x \times N_y \times N_z=1024\times512\times513$ grid points; the computational grid has uniform spacing in the homogenous directions and Chebyshev-Gauss-Lobatto points in the wall-normal direction.

The flow is driven by an imposed constant pressure gradient in the streamwise direction; the shear Reynolds number is kept fixed at $Re_\tau=300$ for all the cases.
Two different values of the reference surface tension (clean interface), set via the Weber number, have been selected: $We=1.50$ (higher surface tension) and $We=3.00$ (lower surface tension). 
The selected values are characteristics of oil/water mixtures \cite{Than1988}.
For each surface tension value (Weber number), four different elasticity numbers have been tested: $\beta_s=0.50$ (weaker surfactant), $\beta_s=1.0$, $\beta_s=2.0$ and $\beta_s=4.0$ (stronger surfactant).
In addition, two simulations of surfactant-free systems ($We=1.50$ and $We=3.00$) and a single-phase flow simulation have been also performed.

For the phase field, the Cahn number is set to $Ch=0.025$. 
This value is selected based on the grid resolution: at least five grid points are required across the interface to accurately describe the steep gradients present \cite{Soligo2019_jcp}. 
The phase field P\'eclet number has been set according to the scaling $Pe_\phi=1/Ch=40$, to achieve the convergence to the sharp interface limit \cite{YUE2007,Magaletti2013}.
Higher grid resolutions allow to reduce the thickness of the interface (smaller Cahn numbers); the computational cost would, however, increase dramatically (roughly as the third power of the refinement factor).
With the aim of evaluating the influence of the grid resolution (and of the Cahn number) on the results, the case $We=3.00$ and $\beta_s=4.00$ has been rerun on a refined grid (twice the resolution in each direction: $N_x \times N_y \times N_z=2048 \times 1024 \times 1025$) and with a halved Cahn number ($Ch=0.0125$).
A specific discussion on the effect of the grid resolution on the results here presented can be found in section~\ref{igrid}.

For the surfactant, the bulk concentration is set equal to $\psi_b=0.01$ for all cases. 
The surfactant P\'eclet number is set to $Pe_\psi=100$, a typical value for nonionic and anionic surfactants in aqueous solutions \cite{Weinheimer1981}.
The temperature-dependent parameter has been set equal to $Pi = 1.35$ while the surfactant solubility parameter to $E_x = 0.117$.
The effect of the surfactant parameters is not investigated here, as we focus on the effects of the global (set via the Weber number) and local (set via the elasticity number) surface tension modifications. 

At the beginning of each simulation, a regular array of 256 spherical droplets with diameter $d=0.4h$ (corresponding to $d^+=120$ wall units) is initialized in a fully-developed single-phase turbulent channel flow. 
The total volume fraction of the dispersed phase is $\Phi=V_d/(V_c+V_d)\simeq5.4\%$, being $V_d$ and $V_c$ the volume of the dispersed and carrier phase respectively. 
The very same configuration (volume fraction, droplet size and number) was used in previous works \cite{Scarbolo2015,Scarbolo2016}, thus allowing for a direct comparison.
The phase field and the surfactant concentration are initialized with their equilibrium profile, equations (\ref{eq: phi_eq})-(\ref{eq: psi_eq}).
As the array of spherical droplets is suddenly released in a single-phase turbulent flow, turbulent velocity fluctuations strongly perturb the interfacial profile; during this initial coupling phase, mass leakages among the phases may occur. After this initial transient, the mass of each phase keeps constant over time.

While the initial condition chosen for the dispersed phase may seem unphysical, after a short transient, memory of the initial condition is completely lost and the analysis performed at statistically steady-state condition are not affected by the initial condition selected.
Different initial conditions have been tested, for instance, the injection of a thin liquid sheet at the channel center, and the same statistically steady-state results were obtained.
In addition, the selected initial condition has a shorter transient before reaching statistically steady-state results with respect to the other configurations tested.
Thus, to reduce the computational cost of the simulations and to better compare the obtained results with previous works \cite{Scarbolo2015,Scarbolo2016,Roccon2017}, the present initial condition for the dispersed phase has been used.

\begin{ruledtabular}
\begin{table}[h]
\centering
\begin{tabular}{c ccccc    ccccccccc}
\multicolumn{10}{c}{Simulations}\\
\hline
System& $Re_\tau$  &  $We$  &  $Ch$  &   $Pe_\phi$  &  $Pe_\psi$ & $Pi$ &$E_x$&$\psi_b$&$\beta_s$\\
\hline
Single-phase    &300& - &-&-&-&-&-&-&- \\
\hline
Surfactant-free &300&1.50&0.025&40&-&-&-&-&- \\
Surfactant-laden&300&1.50&0.025&40&100&1.35&0.117&0.01&0.50 \\
Surfactant-laden&300&1.50&0.025&40&100&1.35&0.117&0.01&1.00 \\
Surfactant-laden&300&1.50&0.025&40&100&1.35&0.117&0.01&2.00 \\
Surfactant-laden&300&1.50&0.025&40&100&1.35&0.117&0.01&4.00 \\
\hline
Surfactant-free  &300&3.00&0.025&40&-&-&-&-&- \\
Surfactant-laden&300&3.00&0.025&40&100&1.35&0.117&0.01&0.50 \\
Surfactant-laden&300&3.00&0.025&40&100&1.35&0.117&0.01&1.00 \\
Surfactant-laden&300&3.00&0.025&40&100&1.35&0.117&0.01&2.00 \\
Surfactant-laden&300&3.00&0.025&40&100&1.35&0.117&0.01&4.00 \\
\end{tabular}
\caption{Overview of the parameters used in the different simulations performed.
For each Weber number, one surfactant-free and four surfactant-laden systems have been analyzed. 
Surfactant-laden systems consider different elasticity numbers: from $\beta_s=0.50$ (weaker surfactant) up to $\beta_s=4.00$ (stronger surfactant).
In addition, a simulation of a single-phase flow has been performed.}
\label{overview}
\end{table}
\end{ruledtabular}

\section{Results}
\label{sec: results}

We characterize the flow modifications produced by the surfactant-laden droplets both from a global and a local point of view.
First, we focus on the macroscopic flow modifications produced by the presence of a surfactant-laden interface.
In particular, we analyze the wall-normal behavior of the mean streamwise velocity profiles and the root mean square of the streamwise and wall-normal velocity fluctuations.
Then, we move to analyze the local flow modifications produced by the surface tension forces, analyzing first the alignment between interface normal and vorticity vector at the interface, and then the distribution of the flow topology parameter, $Q$, in the two phases (carrier and dispersed) and at the interface.
In this work, we focus only on the flow modifications produced by the droplets; a detailed characterization of the dispersed phase morphology can be found in our previous related work \cite{Soligo2019_jfm}.

\subsection{Macroscopic flow modifications}

We quantify the effects of the dispersed phase on the macroscopic flow, analyzing the changes in the mean streamwise velocity and in the root mean square of the streamwise and wall-normal velocity fluctuations.
Once the droplets are released in the turbulent channel flow, a transient is required to adapt the flow field to the presence of the droplets.
After this transient, the system reaches a new steady-state configuration and the statistics are computed.
The statistics have been obtained averaging the entire velocity fields in the two homogeneous directions and over time.

\subsubsection{Mean velocity profiles}

\begin{figure}
\center
\setlength{\unitlength}{0.0025\columnwidth}
\begin{picture}(400,150)
\put(-20,-25){\includegraphics[width=0.65\columnwidth, keepaspectratio]{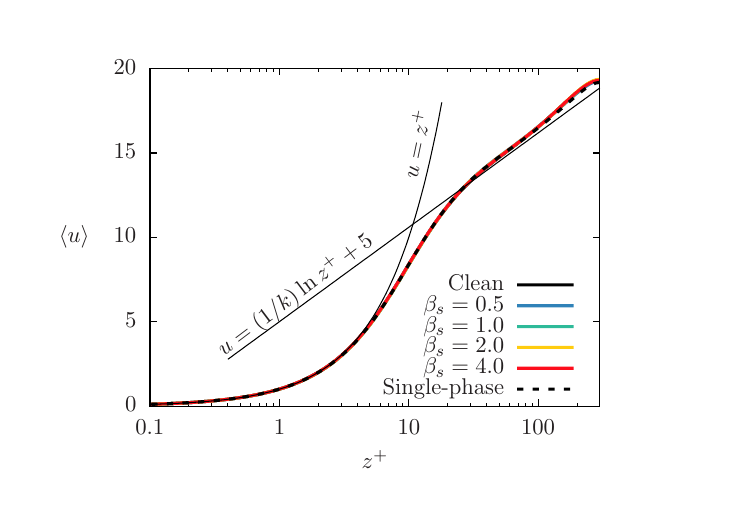}}
\put(180,-25){\includegraphics[width=0.65\columnwidth, keepaspectratio]{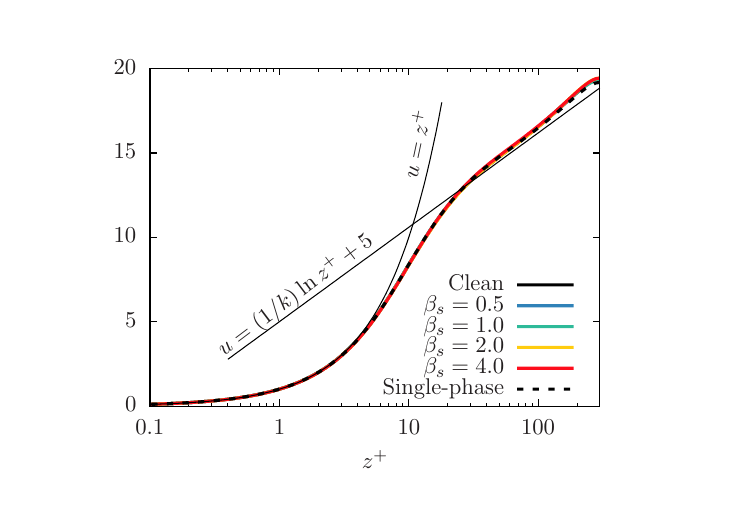}}
\put(5,120){(a)}
\put(200,120){(b)}
\put(84,140){$We=1.50$}
\put(282,140){$We=3.00$}
\end{picture}
\caption{Profiles of the mean streamwise velocity, $\langle u \rangle$ at $Re_\tau=300$ for $We=1.50$ and $We=3.00$ ($z^+$ axis is reported in log scale, base 10).
The single-phase velocity profile is reported with a black dashed line while the multiphase cases are reported with different colors: $\beta_s=0.50$ (blue), $\beta_s=1.00$ (green), $\beta_s=2.00$ (yellow) and $\beta_s=4.00$ (red).
For all multiphase cases, mean velocity profiles are overlapped and a small increase of the velocity profiles can be observed with respect to the single-phase.
The classical law of the wall, $u=z^+$ and $u=(1/k) \ln z^+ +5$ (being $k=0.41$ the von K{\'a}rm{\'a}n constant), is also reported as a reference. }
\label{fig: u_mean}
\end{figure}
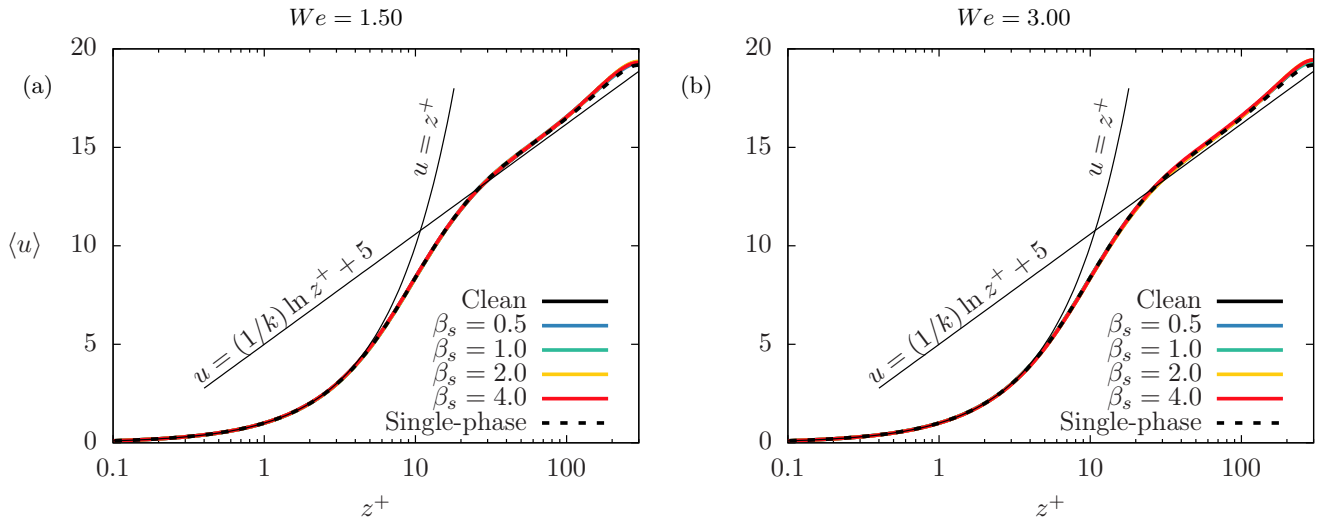

We start by considering the profiles of the mean streamwise velocity, $\langle u \rangle$, figure~\ref{fig: u_mean}.
Panel (a) refers to $We=1.50$, while panel (b) refers to $We=3.00$.
The multiphase cases are reported with solid lines using different colors: clean (black), $\beta_s=0.50$ (blue), $\beta_s=1.00$ (green), $\beta_s=2.00$ (yellow) and $\beta_s=4.00$ (red).
The single-phase results are indicated using a dashed black line.
In addition, the classic law of the wall \cite{schlichting2016boundary}, $u=z^+$ and $u=(1/k) \ln z^+ +5$ (where $k=0.41$ is the von K{\'a}rm{\'a}n constant \cite{von1931mechanical}), is reported as reference.
For both Weber numbers, we can observe that all the mean streamwise velocity profiles are almost unaffected by the presence of the dispersed phase and all multiphase cases fall almost on top of each other.
Thus, for the range of parameters here considered, the surface tension modifications, either global (Weber number) either local (elasticity number), do not modify significantly the mean flow.
Nevertheless, with respect to the single-phase case (black dashed), we can observe a slight increase of the mean velocity and indeed an increase in the flow-rate of about 1\% is observed for all the multiphase cases.
These observations are in agreement with the results obtained previously by Scarbolo {\em et al.} 2016 \cite{Scarbolo2016} and Roccon {\em et al.} 2017 \cite{Roccon2017}, in a similar configuration, albeit at a lower Reynolds number ($Re_\tau=150$).
In this paper, we performed new simulations at $Re_\tau=150$ maintaining the grid resolution, so to better address the effect of the Reynolds number on the macroscopic flow indicators; a full discussion on this issue can be found in appendix~\ref{appa}.

\subsubsection{Root mean square (RMS) of the velocity fluctuations}

\begin{figure}
\center
\setlength{\unitlength}{0.0025\columnwidth}
\begin{picture}(400,350)
\put(-20,130){\includegraphics[width=0.65\columnwidth, keepaspectratio]{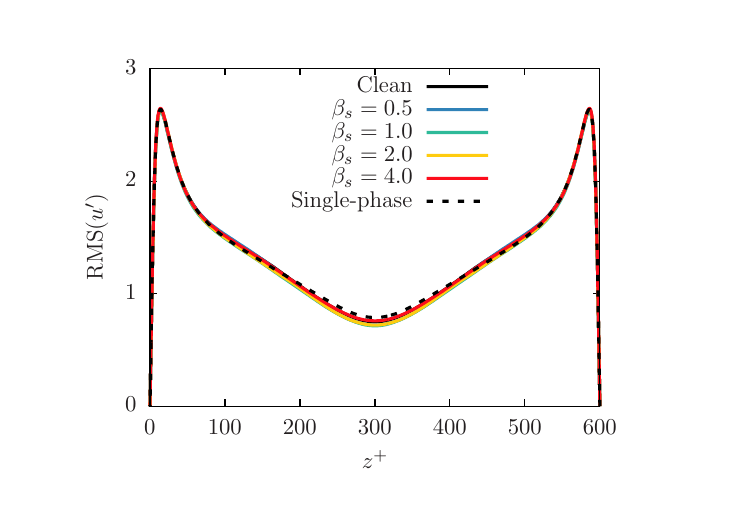}}
\put(180,130){\includegraphics[width=0.65\columnwidth, keepaspectratio]{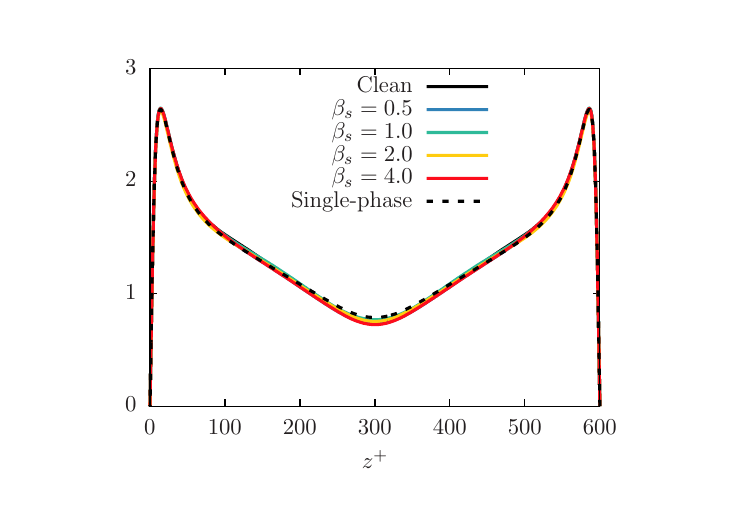}}
\put(-20,-25){\includegraphics[width=0.65\columnwidth, keepaspectratio]{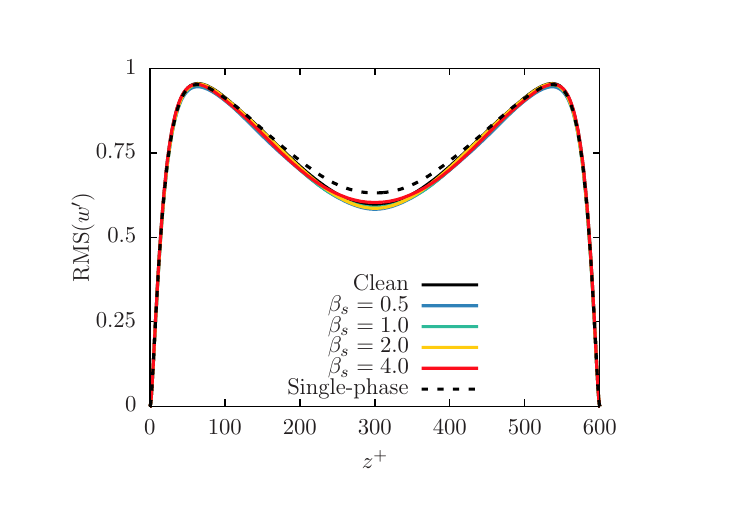}}
\put(180,-25){\includegraphics[width=0.65\columnwidth, keepaspectratio]{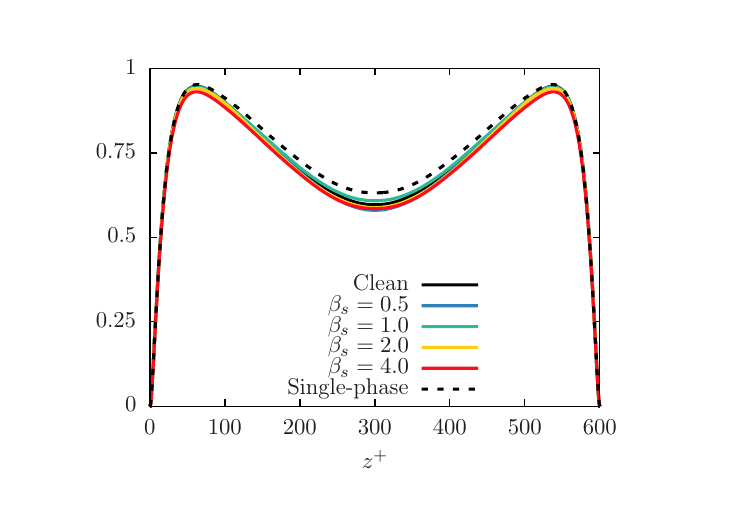}}
\put(5,275){(a)}
\put(200,275){(b)}
\put(5,120){(c)}
\put(200,120){(d)}
\put(88,295){$We=1.50$}
\put(286,295){$We=3.00$}
\end{picture}
\caption{Root mean square (RMS) of the velocity fluctuations, panels (a)-(b) for the streamwise component $u'$ and panels (c)-(d) for the wall-normal component $w'$. The left column, panels (a)-(c), refers to $We=1.50$, while the right column, panels (b)-(d), refers to $We=3.00$. 
The different cases are reported with different colors: clean (black), $\beta_s=0.50$ (blue), $\beta_s=1.00$ (green), $\beta_s=2.00$ (yellow) and $\beta_s=4.00$ (red).
A black dashed line identifies single-phase flow results.
For all multiphase cases, a slight reduction of the amplitude of the velocity fluctuations is observed in the channel centre.}
\label{fig: uw_rms}
\end{figure}
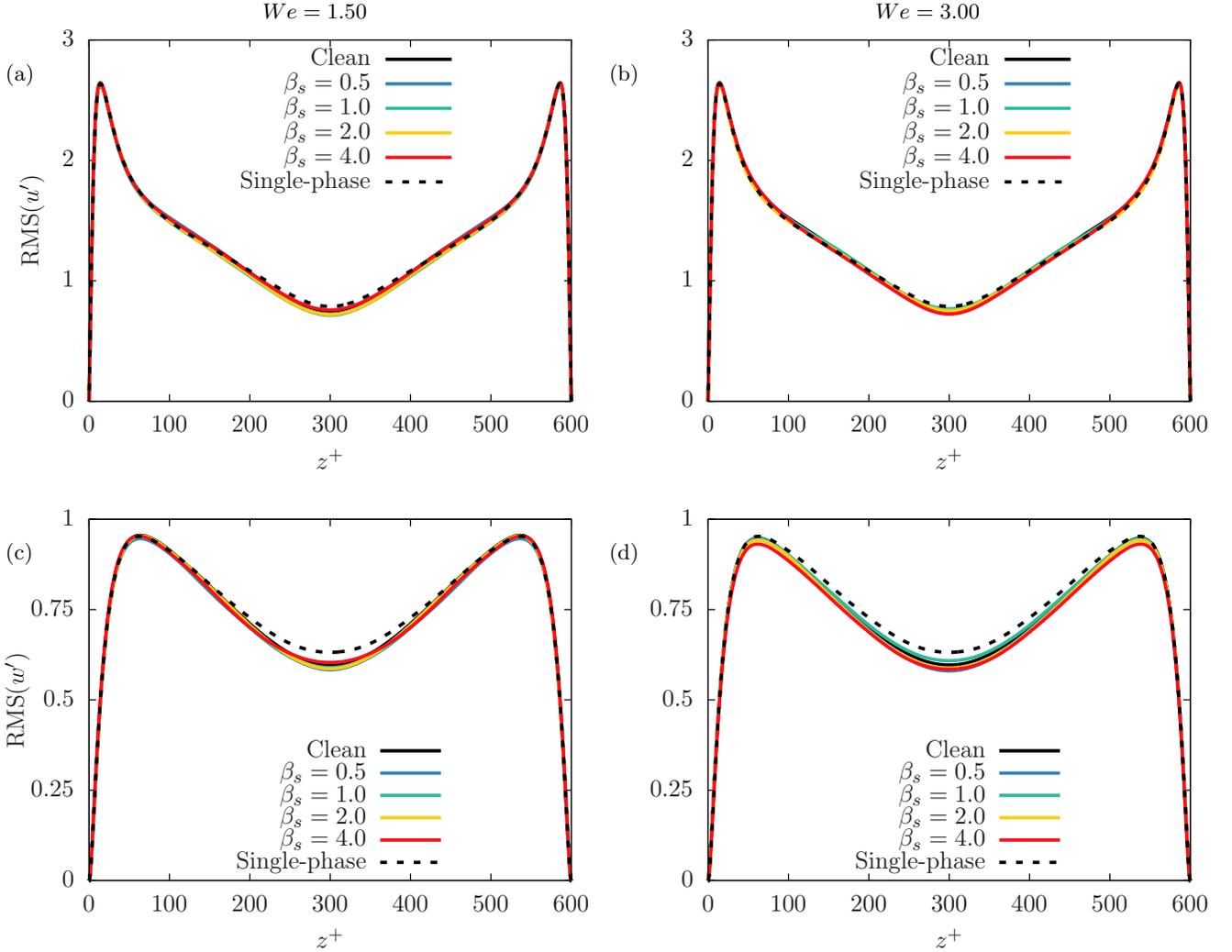

More significative effects of the dispersed phase on the flow statistics can be appreciated in the root mean square (RMS) of the streamwise and wall-normal velocity fluctuations, which are shown in figure~\ref{fig: uw_rms}.
The left column, panels (a)-(c), refers to $We=1.50$, while the right column, panels (b)-(d), refers to $We=3.00$.
The multiphase cases are reported with different colors: clean (black), $\beta_s=0.50$ (blue), $\beta_s=1.00$ (green), $\beta_s=2.00$ (yellow) and $\beta_s=4.00$ (red), while the single-phase results are reported with a black dashed line.
We focus on the modifications which occur in the core of the channel, where most of the droplets are located.
For all the multiphase cases, with respect to the single-phase, we observe a decrease in the amplitude of the streamwise and wall-normal velocity fluctuations.
This decrease is more pronounced for the wall-normal velocity fluctuations (figure~\ref{fig: uw_rms}c-d), while it is milder for the streamwise velocity fluctuations (figure~\ref{fig: uw_rms}a-b).
These modifications can be directly traced back to the presence of the droplets: indeed, the interface acts as a decoupling element between the carrier phase fluid and the dispersed phase fluid, modulating turbulence and momentum exchanges across it.
Thus, the presence of the interface damps the fluid velocity fluctuations (in all directions).
There is also an additional factor, which will be better addressed and quantified later on, which modifies the local flow within the droplets: the confinement effect due to the interface enclosing portions of fluid.
Specifically, the fluid within the droplets is confined by the interface, which modulates momentum exchanges with the external flow and suppresses the larger flow structures enclosed by the interface.
This confinement effect is inversely proportional to the size of droplets: internal flowing is strongly suppressed in smaller droplets. 
This latter aspect will be better addressed in the section dedicated to the flow topology parameter, in which the effect of the flow confinement will be clear (see figure~\ref{fig: Q_drop} and the related discussion).

\subsection{Local flow modifications}

We move now to analyze the modifications produced by the surface tension forces (capillary and Marangoni) on the local flow field.
First, we consider the changes in the local vorticity at the interface \cite{shao2018direct,mukherjee2019}.
Then, we consider the spatial distribution of the flow topology parameter \cite{De2017_nonnew,Devita2019} in the different regions of the domain: carrier phase, dispersed phase and at the interface. 

\subsubsection{Vorticity and interface normal alignment}

\begin{figure}
\center
\setlength{\unitlength}{0.0025\columnwidth}
\begin{picture}(400,150)
\put(88,140){$We=1.50$}
\put(286,140){$We=3.00$}
\put(-20,-25){\includegraphics[width=0.65\columnwidth, keepaspectratio]{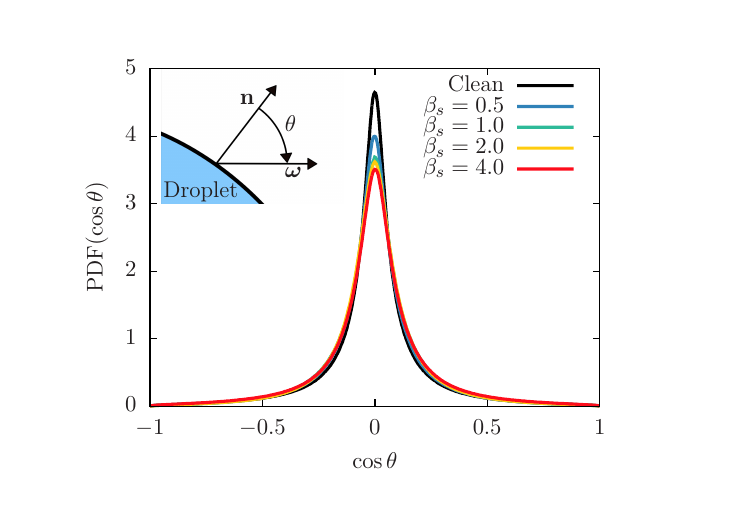}}
\put(180,-25){\includegraphics[width=0.65\columnwidth, keepaspectratio]{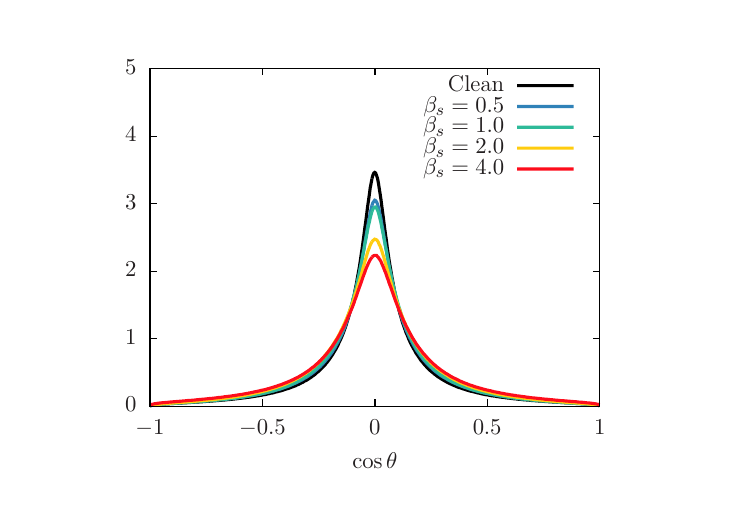}}
\put(20,120){(a)}
\put(206,120){(b)}
\end{picture}
\caption{Probability density function (PDF) of  $\cos\theta$ computed at the interface, being $\theta$ the angle between normalized vorticity $\boldsymbol{\omega}/|\boldsymbol{\omega}|$ and interface normal $\mathbf{n}$ (see inset).
Panel (a) refers to $We=1.50$, while panel (b) to $We=3.00$. 
The different cases are reported using different colors: clean (black), $\beta_s=0.50$ (blue), $\beta_s=1.00$ (green), $\beta_s=2.00$ (yellow) and $\beta_s=4.00$ (red).
We can distinguish among three different reference configurations: $\cos \theta=-1$ (interface normal and vorticity opposed), $\cos \theta=0$ (interface normal and vorticity perpendicular) and $\cos \theta=+1$ (interface normal and vorticity concordant).}
\label{fig: vort}
\end{figure}
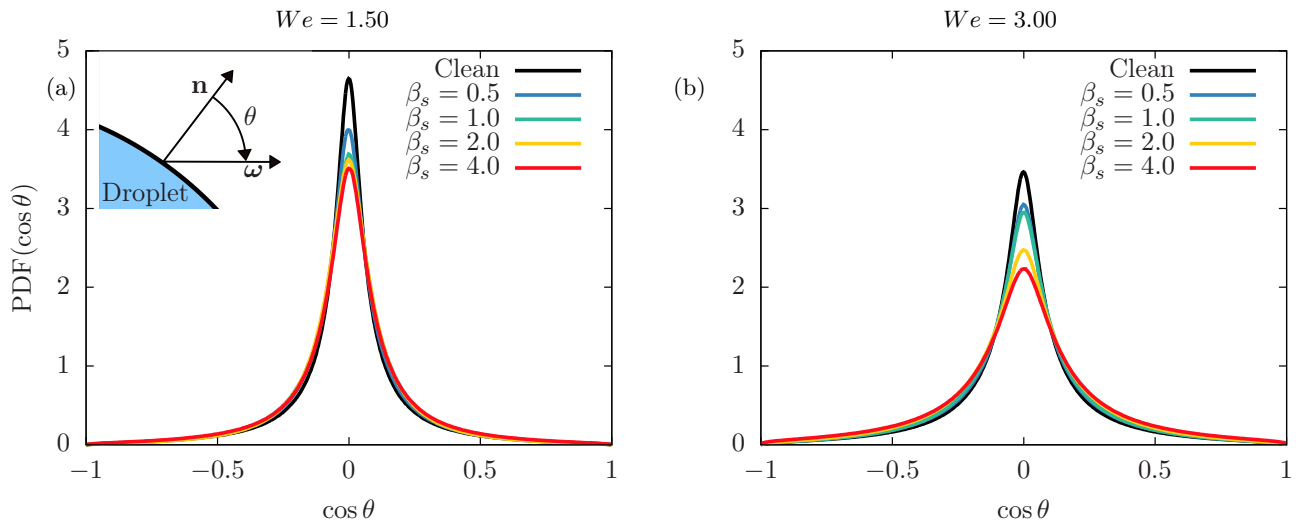

To characterize the modifications produced by the capillary and Marangoni forces in the vorticity field at the interface, we analyze the alignment between the vorticity vector, $\boldsymbol{\omega}=\nabla\times \mathbf{u}$, and the interface normal, $\mathbf{n}$.
In particular, we compute the cosine of the angle $\theta$, where $\theta$ is the angle between the vorticity vector (evaluated at the interface) and the interface normal (see sketch of figure~\ref{fig: vort}).
Using a phase-field approach, the interface normal can be directly computed from the phase field \cite{Aris1989,Sun2007,Roccon2017}:
\begin{equation}
\mathbf{n}=-\frac{\nabla\phi}{|\nabla\phi|} \; ,
\end{equation}
where the minus sign is needed to obtain the outward-pointing normal (from the droplet towards the carrier phase, see the sketch in figure~\ref{fig: vort}).
To obtain the cosine of the angle $\theta$, we can compute the scalar product between the interface normal and the vorticity normalized by its magnitude: 
\begin{equation}
\cos\theta=\mathbf{n}\cdot \frac{\boldsymbol{\omega}}{|\boldsymbol{\omega}|} \; .
\end{equation}
The range of values assumed by the cosine of $\theta$ spans from  $\cos \theta = -1$ (interface normal and vorticity opposed), to $\cos \theta = 0$ (interface normal and vorticity perpendicular) up to 
$\cos \theta = +1$ (interface normal and vorticity concordant).

To characterize the alignment between these two quantities at the interface of the droplets, we compute the probability density function (PDF) of $\cos\theta$.
The results are shown in figure~\ref{fig: vort}; panel (a) refers to $We=1.50$, while panel (b) to $We=3.00$.
The different cases are reported using different colors: clean (black), $\beta_s=0.50$ (blue), $\beta_s=1.00$ (green), $\beta_s=2.00$ (yellow) and $\beta_s=0.50$ (red).

We start by considering the cases at $We=1.50$ (figure~\ref{fig: vort}a).
For all cases, we observe that the PDF is symmetric and the most probable value is $\cos\theta=0$.
By opposite, larger values of $\cos \theta$ (in magnitude) are very unlikely to be found.
This indicates that there is a strong probability that the interface normal and the vorticity vector are perpendicular.
Conversely, there is a much lower probability that the interface normal and the vorticity are aligned along the same direction (with same or opposite sign).
The larger probability of finding vorticity tangential to the interface can be addressed to the action of the surface tension forces \cite{leal1992,batchelor2000}. 
Increasing the elasticity number (from clean, i.e. $\beta_s=0$, to $\beta_s=4.00$), we can appreciate a clear modification of the PDF shape, which becomes slightly wider and the amplitude of the peak shifts downwards (from $\simeq 4.6$ for the surfactant-free system down to $\simeq 3.5$ for $\beta_s=4.00$).
This modification of the PDF is produced by the action of the surfactant, which reduces the local surface tension (and also the average).
Thus, the interface becomes more deformable and in turn less capable of modifying the local vorticity field.

Moving to $We=3.00$ (figure~\ref{fig: vort}b), a similar trend can be appreciated.   
However, with respect to $We=1.50$, the PDFs are slightly wider and the amplitude of the peaks decreases. 
Also for these cases, the widening of the PDFs can be traced back to the lower surface tension. 
Indeed, when higher Weber numbers (i.e. lower reference surface tension) or stronger surfactants (i.e. larger elasticity numbers) are considered, the PDFs become wider and the amplitude of the peak further decreases (from $\simeq 3.5$ for the surfactant-free system down to $\simeq 2.1$ for $\beta_s=4.00$).

Finally, it is interesting to observe that the PDFs of the cases $We=1.50$ and $\beta_s=4.00$ (figure~\ref{fig: vort}a, red) and the clean case at $We=3.00$ (figure~\ref{fig: vort}b, black) are characterized by a similar behavior.
This similarity can be justified considering that the average surface tension of the case $We=1.50$ and $\beta_s=4.00$ is $\sigma_{av} \simeq 0.55 \sigma_0$.
Thus, its equivalent Weber number, $We_{eq}= ( \sigma_0/\sigma_{av}) We$, computed using the average surface tension as a reference is $We_{eq} \simeq 2.7$.
As a consequence, the PDFs of the two cases are very similar.

In an effort to investigate the effect of the Reynolds number on the vorticity and interface normal alignment, we have examined the same statistics for a lower Reynolds number running two additional simulations.
A detailed discussion is reported in appendix~\ref{appa}.

\subsubsection{Flow topology parameter}  

\begin{figure}
\setlength{\unitlength}{0.0025\columnwidth}
\begin{picture}(400,110)
\put(20,10){\includegraphics[width=0.9\columnwidth, keepaspectratio]{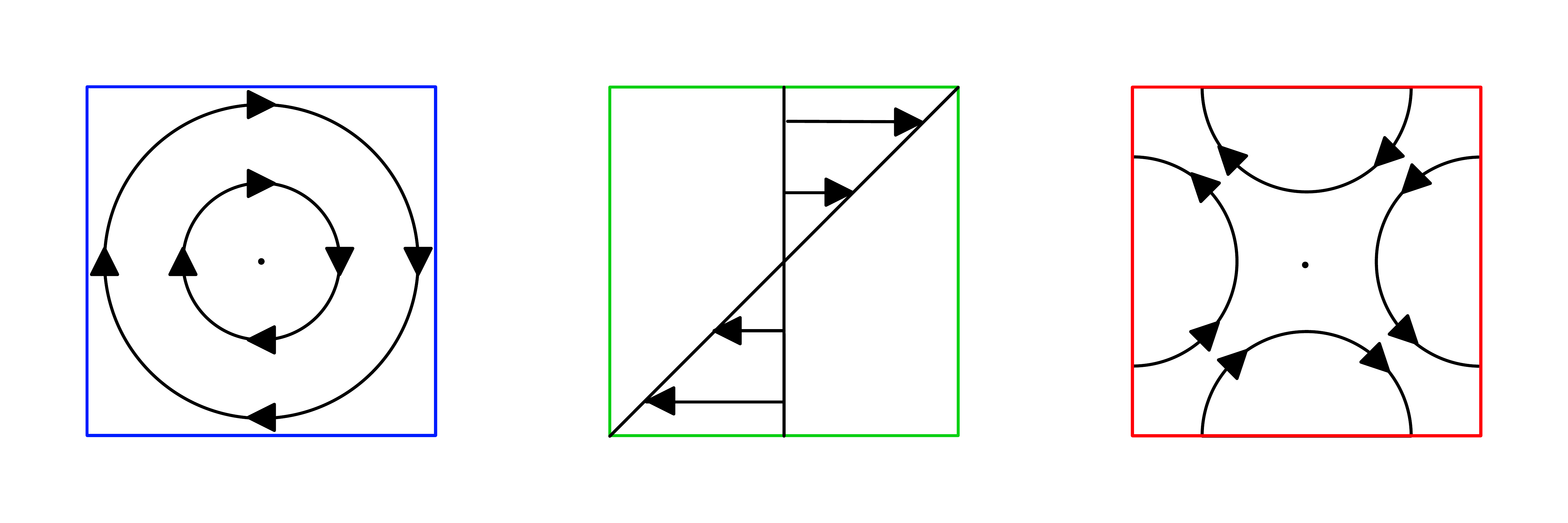}}
\put(63,15){$\Q=-1$}
\put(185,15){$\Q=0$}
\put(302,15){$\Q=+1$}
\put(57,0){Rotational}
\put(175,0){Pure shear}
\put(292,0){Elongational}
\put(25,100){(a)}
\put(145,100){(b)}
\put(265,100){(c)}
\end{picture}
\caption{Qualitative representation of the flow topologies obtained for different values of $\Q$.
Panel (a) shows a rotational flow ($\Q=-1$), panel (b) shows a pure shear flow ($\Q=0$) and panel (c) shows an elongational flow ($\Q=+1$).
The boxes are colored by the value of $\Q$.} 
\label{qsketch}
\end{figure}
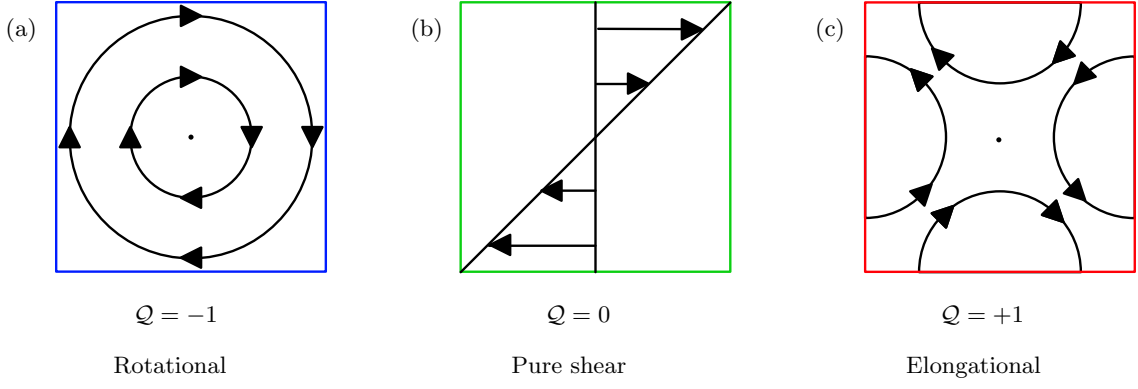

The flow topology parameter $\Q$ \cite{Perry1987} is used to characterize the effects of the interface on the flow field in the different regions of the domain (carrier phase, dispersed phase and interface).
The parameter $\Q$  recently gained popularity in the study of multiphase flows \cite{De2017_nonnew,Devita2019,RostiDB_2019,Devita2019_nonnew,dodd2019small} and allows  to distinguish among three different type of flow topology: purely rotational flow ($\Q=-1$, figure~\ref{qsketch}a), pure shear flow  ($\Q=0$, figure~\ref{qsketch}b) and purely elongational flow ($\Q=+1$, figure~\ref{qsketch}c).
The flow topology parameter $\Q$ is calculated as the second invariant of the velocity gradient tensor, $\nabla \mathbf{u}$:
\begin{equation}
\Q=\frac{D^2-\Omega^2}{D^2+\Omega^2}=
\begin{cases}
-1   &~~\textnormal{Purely rotational flow}\\
~~0   &~~\textnormal{Pure shear flow} \\
+1    &~~\textnormal{Purely elongational flow} \\
\end{cases}
\; ,
\end{equation}
with $D^2=\mathsf{D}:\mathsf{D}$ and $\Omega^2=\mathsf{\Omega}:\mathsf{\Omega}$, where $\mathsf{D}$ and $\mathsf{\Omega}$ are the rate-of-deformation and rate-of-rotation tensors \cite{batchelor2000} respectively and are defined as follows:
\begin{equation}
\mathsf{D}=\frac{\nabla \mathbf{u} +\nabla \mathbf{u}^T}{2} \, ,
\end{equation}
\begin{equation}
\mathsf{\Omega}=\frac{\nabla \mathbf{u} -\nabla \mathbf{u}^T}{2} \, .
\end{equation}

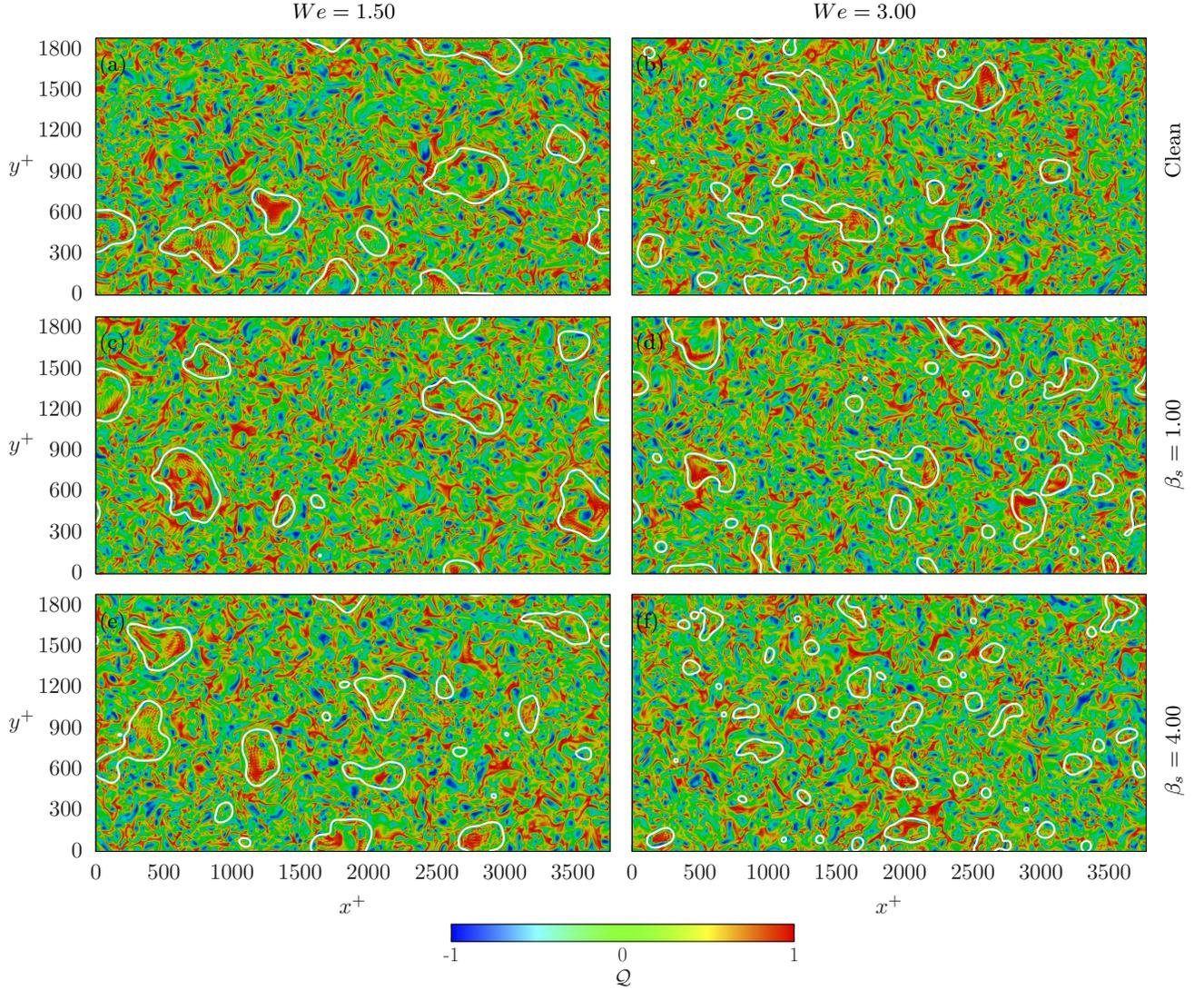
\begin{figure}
\setlength{\unitlength}{0.0025\columnwidth}
\begin{picture}(400,315)
\put(0,181){\includegraphics[width=0.6\columnwidth, keepaspectratio]{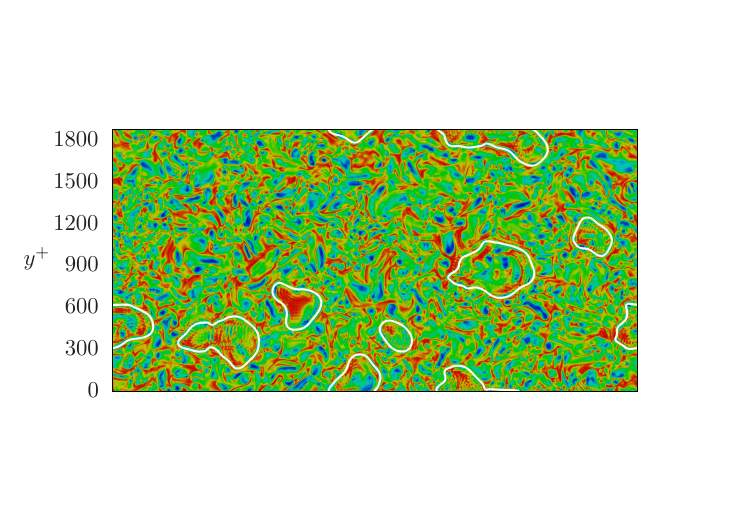}}
\put(175,181){\includegraphics[width=0.6\columnwidth, keepaspectratio]{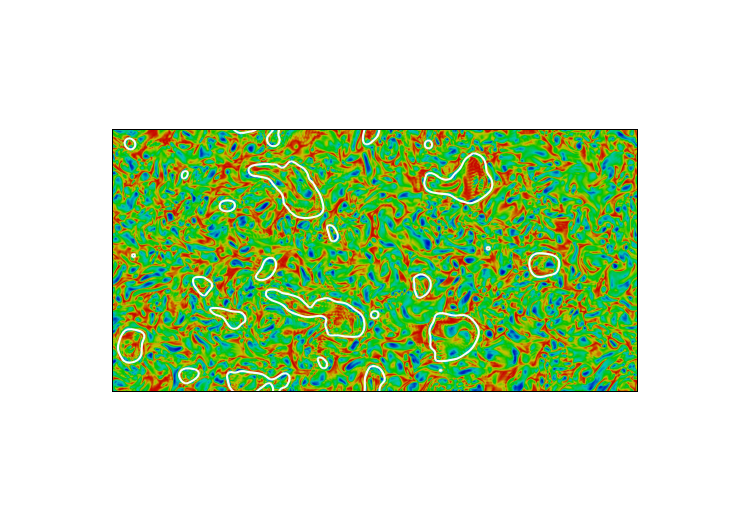}}
\put(0,90){\includegraphics[width=0.6\columnwidth, keepaspectratio]{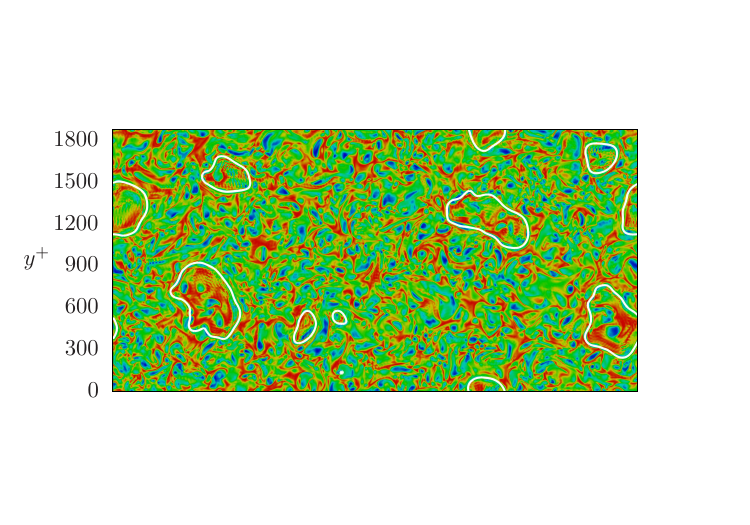}}
\put(175,90){\includegraphics[width=0.6\columnwidth, keepaspectratio]{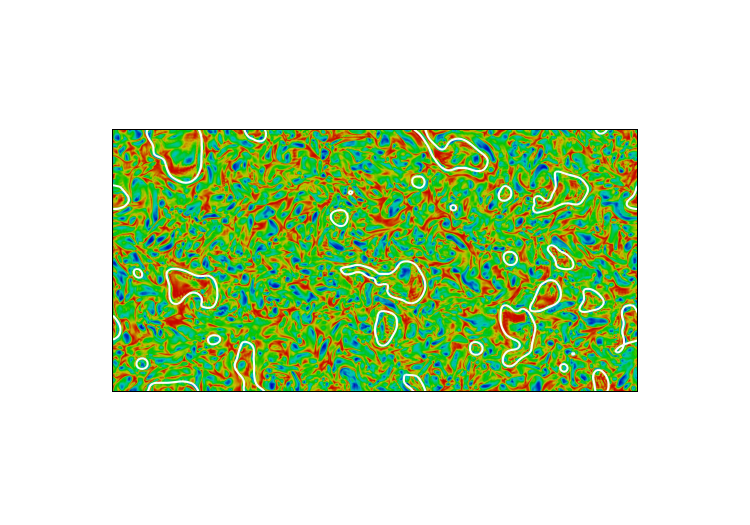}}
\put(0,-8){\includegraphics[width=0.6\columnwidth, keepaspectratio]{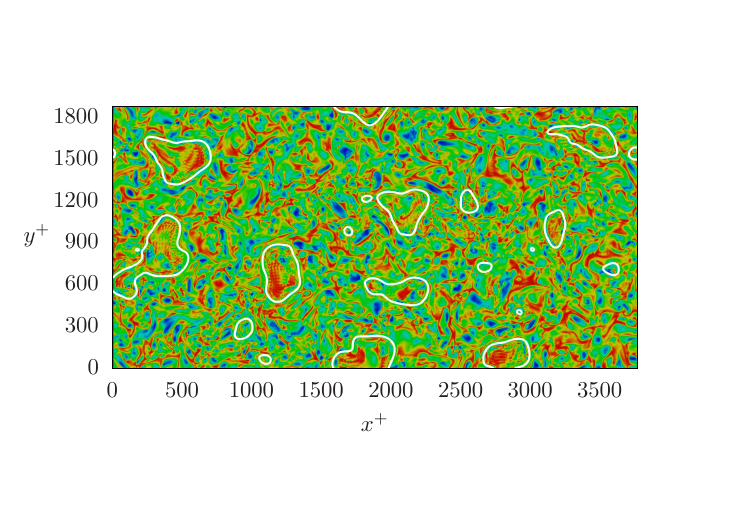}}
\put(175,-8){\includegraphics[width=0.6\columnwidth, keepaspectratio]{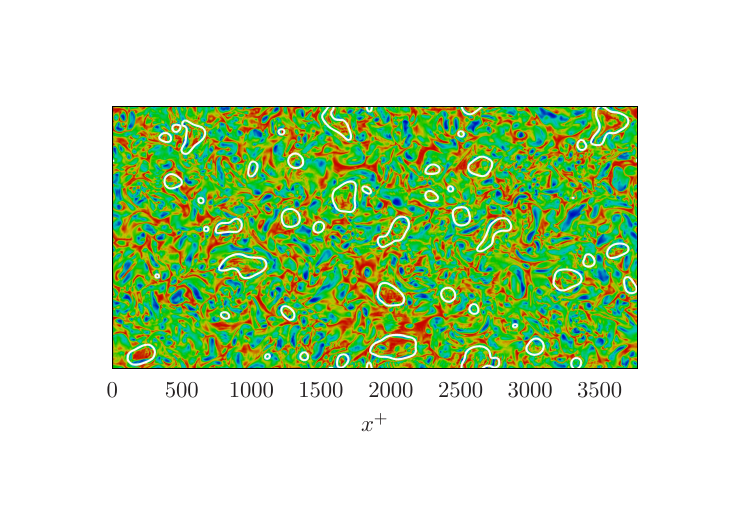}}
\put(138,-5){\includegraphics[width=0.35\columnwidth, keepaspectratio]{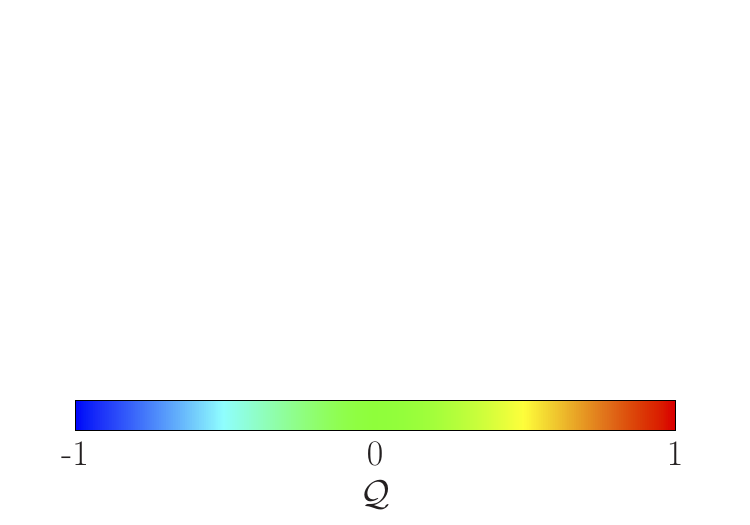}}
\put(100,314){$We=1.50$}
\put(270,314){$We=3.00$}
\put(385,262){\rotatebox{90}{Clean}}
\put(385,160){\rotatebox{90}{$\beta_s=1.00$}}
\put(385,60){\rotatebox{90}{$\beta_s=4.00$}}
\put(37,297){(a)}
\put(212,297){(b)}
\put(37,206){({c})}
\put(212,206){({d})}
\put(37,115){({e})}
\put(212,115){({f})}
\end{picture}
\caption{Flow topology parameter $\Q$ for different Weber and elasticity numbers at $t^+=3750$. 
Top row, panels (a)-(b), refers to surfactant-free simulations; middle row, panels (c)-(d), to $\beta_s=1.0$ (mild surfactant) and bottom row, panels (e)-(f), to $\beta_s=4.00$ (stronger surfactant). 
The left column, panels (a)-(c)-(e), refers to $We=1.50$, while the right column, panels (b)-(d)-(f), to $We=3.00$. 
Each panel shows the flow topology parameter $\Q$ on the channel mid-plane ($x^+-y^+$ plane); white solid lines identify the position of the droplets interface (iso-level $\phi=0$).}
\label{fig: quali}
\end{figure}

We start discussing the behavior of the flow topology parameter $\Q$ qualitatively.
Figure~\ref{fig: quali} shows the spatial distribution of the flow topology parameter $\Q$ in a $x^+-y^+$ plane located at the channel center for six different cases.
Panels (a)-(c)-(e) refer to $We=1.50$, while panels (b)-(d)-(f) refer to $We = 3.00$. 
Each row corresponds to a different elasticity number: surfactant-free (panels~a-b), $\beta_s=1.00$ (panels~c-d) and $\beta_s=4.00$ (panels~e-f). 
The interface of the droplets (iso-level $\phi=0$) is reported using solid white lines.
At a first glance, we can observe how the presence of the interface influences the local flow behavior: while the carrier phase appears to be characterized mainly by pure shear flow regions ($\Q=0$, green) with small fragmented regions of rotational ($\Q=-1$, blue) and elongational ($\Q=+1$, red) flow, the dispersed phase seems to be characterized by a strong presence of both shear and elongational flow regions.
From figure~\ref{fig: quali} we can also make some considerations on the effects of the Weber and elasticity numbers on the dispersed phase morphology, which will be useful later on.
In particular, for increasing Weber and/or elasticity numbers, the number of droplets increases and, as a consequence, the average size of the droplets decreases (constant volume fraction).
This modification of the dispersed phase morphology can be directly related to the lower surface tension corresponding to larger Weber and/or elasticity numbers \cite{Soligo2019_jfm}.
To quantify these observations, we compute the probability density function (PDF) of $\Q$ distinguishing among the different regions of the domain: carrier phase, dispersed phase and interface.

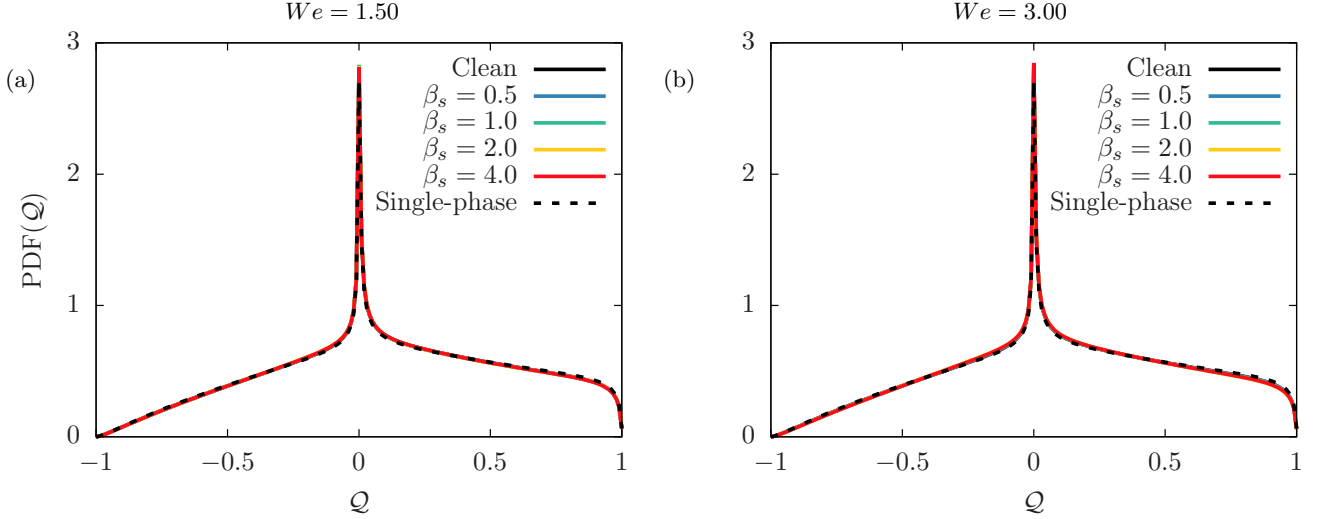
\begin{figure}
\center
\setlength{\unitlength}{0.0025\columnwidth}
\begin{picture}(400,150)
\put(-20,-25){\includegraphics[width=0.65\columnwidth, keepaspectratio]{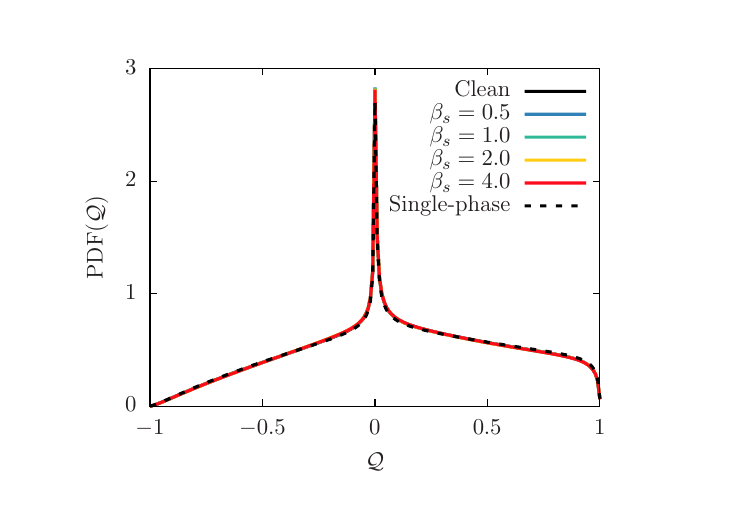}}
\put(180,-25){\includegraphics[width=0.65\columnwidth, keepaspectratio]{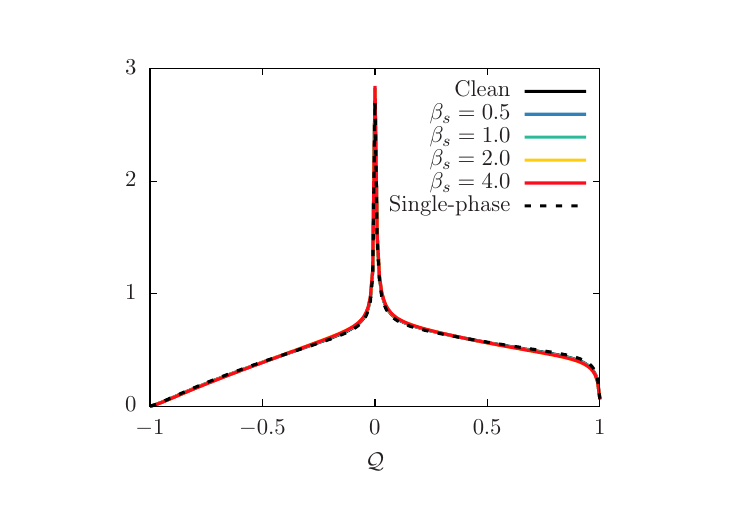}}
\put(5,120){(a)}
\put(200,120){(b)}
\put(88,140){$We=1.50$}
\put(286,140){$We=3.00$}
\end{picture}
\caption{Probability density function (PDF) of the flow topology parameter $\Q$ in the carrier phase.
Panel (a) refers to the cases at $We=1.50$, while panel (b) to the cases at $We=3.00$. 
The different cases are reported using different colors: clean (black), $\beta_s=0.50$ (blue), $\beta_s=1.00$ (green), $\beta_s=2.00$ (yellow) and $\beta_s=4.00$ (red).
The single-phase flow results  (black dashed line) are also reported as a reference.
In the carrier phase, no significative differences are observed and all the multiphase cases overlap with the single-phase flow results.}
\label{fig: Q_carrier}
\end{figure}

In figure~\ref{fig: Q_carrier} we show the PDF of $\Q$ in the carrier phase; panel (a) refers to $We=1.50$, while panel (b) to $We=3.00$. 
The different cases are reported using different colors: single-phase (black dashed), surfactant-free (black), $\beta_s=0.50$ (blue), $\beta_s=1.00$ (green), $\beta_s=2.00$ (yellow) and $\beta_s=4.00$ (red).
For the single-phase case, the PDF is slightly asymmetric and positive values of $\Q$ (elongational flow events) are more likely to be found (with respect to negative values).
In addition, the PDF exhibits a very sharp peak for $\Q=0$ (pure shear flow events) indicating that the single-phase flow is dominated by pure shear flow events and that purely rotational and purely elongational flow events are less frequent.
For the multiphase cases (surfactant-free and surfactant-laden), the situation is unchanged and all the PDFs overlap with the single-phase flow results. 
This suggests that the low volume fraction strongly limits the modifications induced by the droplets in the flow topology of the carrier \cite{RostiGJDB_2019}.
Thus, the mean shear generated by the two walls bounding the flow determines the flow topology in the carrier (which is identical to that observed in a canonical single-phase flow).

\begin{figure}
\center
\setlength{\unitlength}{0.0025\columnwidth}
\begin{picture}(400,150)
\put(-20,-25){\includegraphics[width=0.65\columnwidth, keepaspectratio]{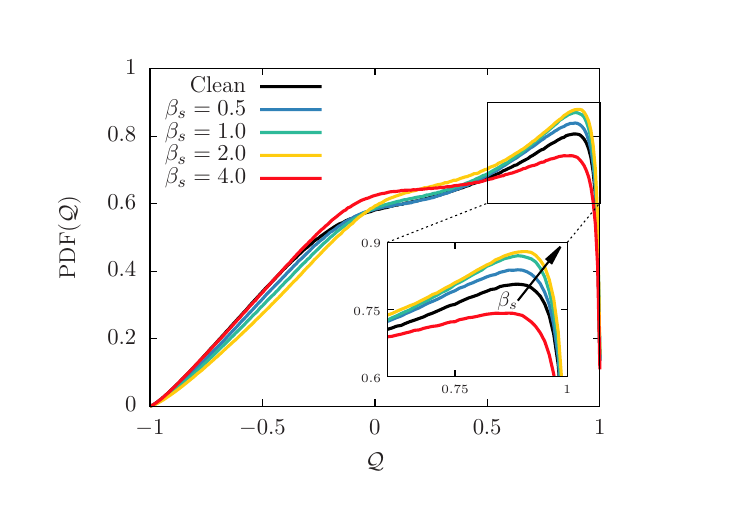}}
\put(180,-25){\includegraphics[width=0.65\columnwidth, keepaspectratio]{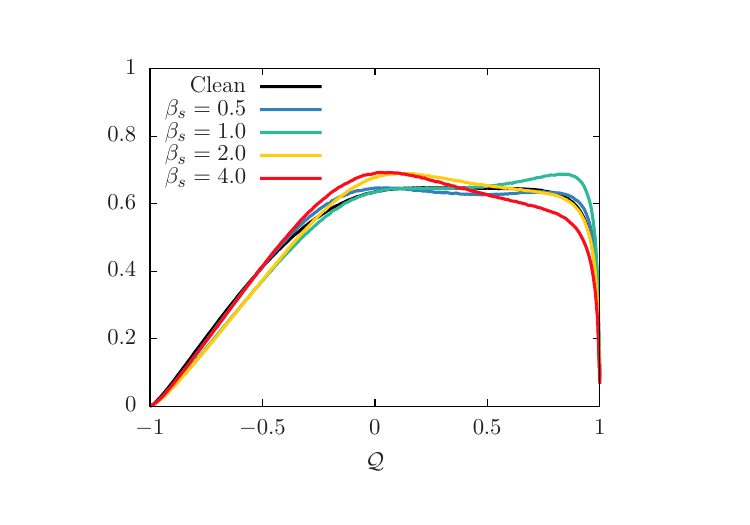}}
\put(5,120){(a)}
\put(200,120){(b)}
\put(88,140){$We=1.50$}
\put(286,140){$We=3.00$}
\end{picture}
\caption{Probability density function (PDF) of the flow topology parameter $\Q$ in the dispersed phase.
Panel (a) refers to $We=1.50$, while panel (b) refers to $We=3.00$. 
The different cases are reported using different colors: clean (black), $\beta_s=0.50$ (blue), $\beta_s=1.00$ (green), $\beta_s=2.00$ (yellow) and $\beta_s=4.00$ (red).
For $We=1.50$, Marangoni forces lead to a larger probability of finding elongational flow events; for $We=3.00$, the lower surface tension and the flow confinement increase the probability of observing pure shear flow events.}
\label{fig: Q_drop}
\end{figure}
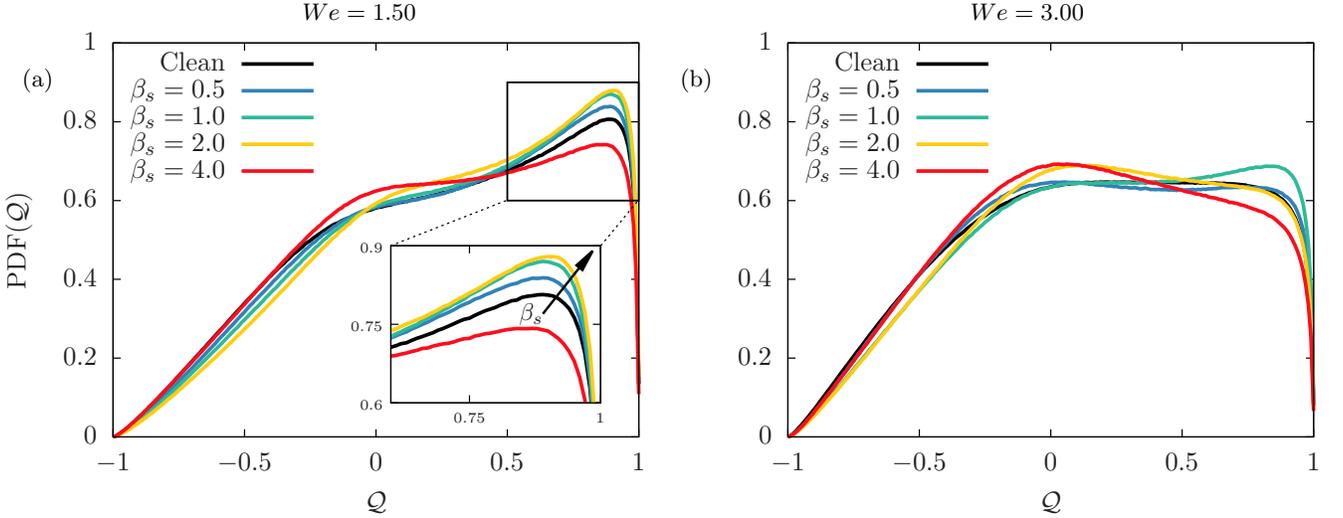

The PDFs of $\Q$ in the dispersed phase are reported in figure~\ref{fig: Q_drop}.
The cases at $We=1.50$ are displayed in panel (a), while those at $We=3.00$ in  panel (b). The different cases are reported using different colors: clean (black), $\beta_s=0.50$ (blue), $\beta_s=1.00$ (green), $\beta_s=2.00$ (yellow) and $\beta_s=4.00$ (red).
Compared to the single-phase flow (figure~\ref{fig: Q_carrier}, black dashed line), the shape of the PDFs is very different for all multiphase cases.

\begin{figure}
\begin{picture}(400,200)
\put(-12,130){High $\psi$ (low $\sigma$)}
\put(-12,14){Low $\psi$ (high $\sigma$)}
\put(-12,35){Marangoni forces}
\put(-15,0){\includegraphics[width=0.95\columnwidth,keepaspectratio]{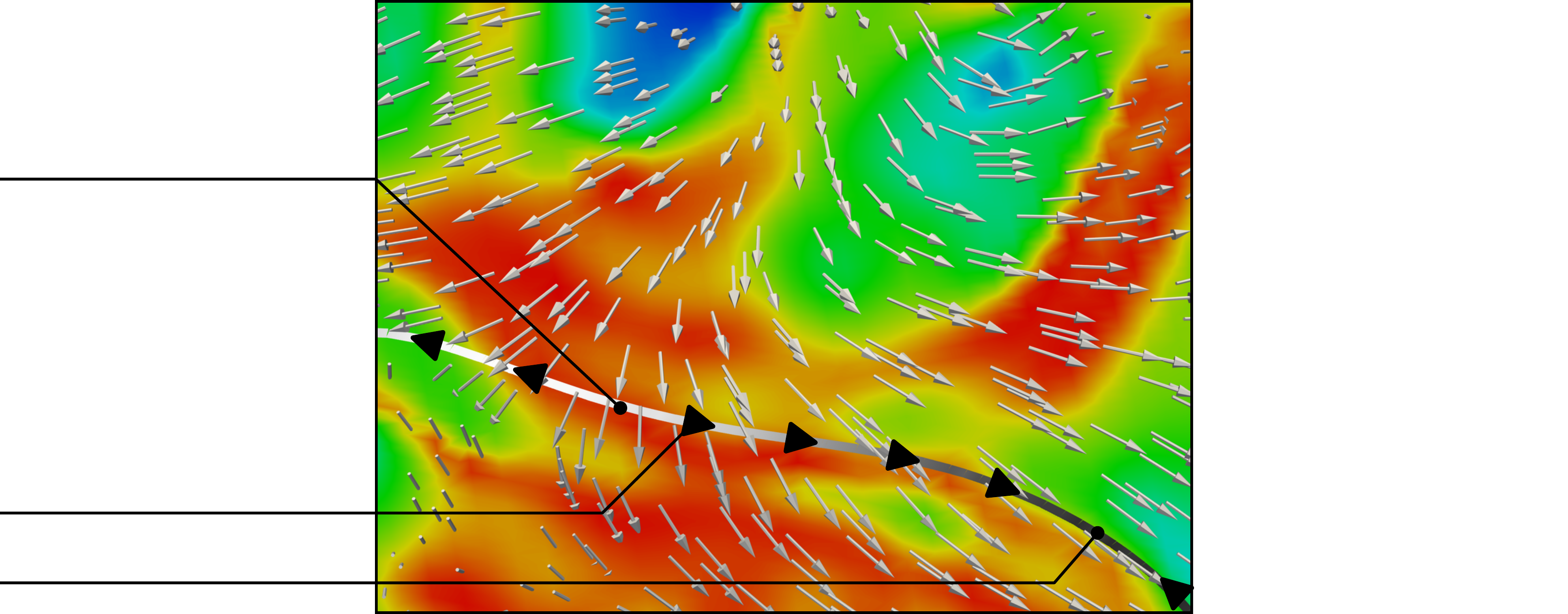}}
\put(255,-12){\includegraphics[width=0.39\columnwidth,keepaspectratio]{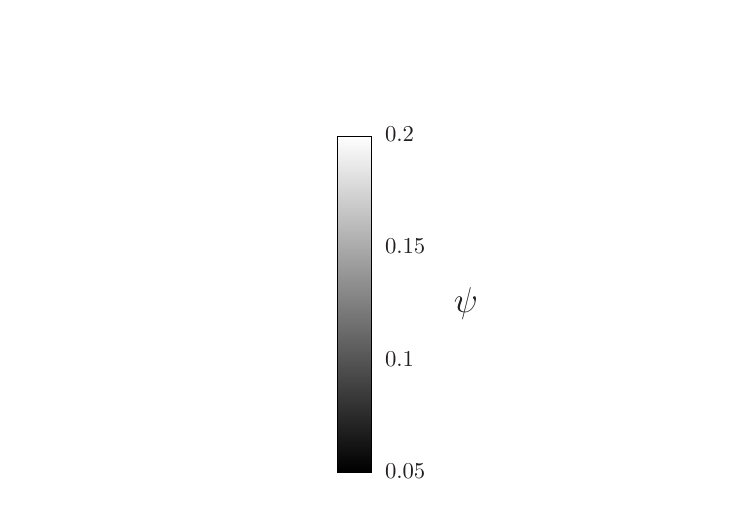}}
\put(255,80){\includegraphics[width=0.39\columnwidth,keepaspectratio]{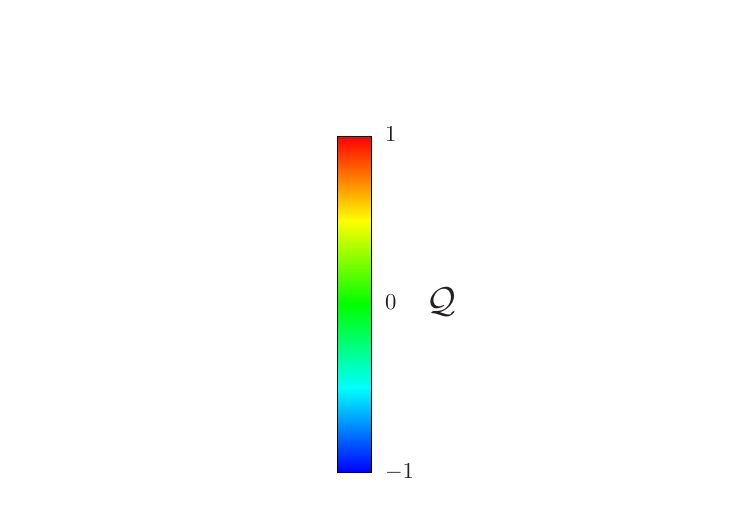}}
\end{picture}
\caption{Qualitative sketch of the elongational-like flow ($\Q=+1$) induced by the Marangoni forces.
The background is colored by the topology parameter $\Q$ (blue-low; red-high) while the interface (iso-level $\phi=0$) is colored by the surfactant concentration (black-low; white-high).
The Marangoni forces (black arrows) are directed from higher surfactant concentration regions (white) towards low surfactant regions (black), thus from regions characterized by a lower surface tension (white) towards regions characterized by a larger surface tension (black).
The background velocity field is represented with gray arrows.}
\label{fig: sketch}
\end{figure}
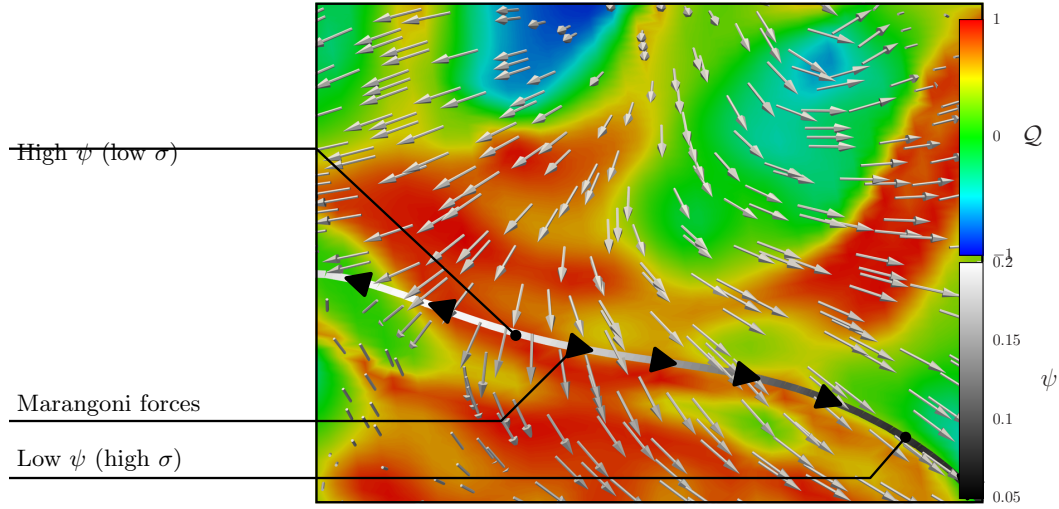

At $We=1.50$ (figure~\ref{fig: Q_drop}a), the PDFs for all cases are negatively skewed and there is a higher probability of observing purely elongational flow events (with respect to pure shear flow events).
We also observe that for larger elasticity numbers (expect the case $\beta_s=4.00$ which will be considered later on), the probability of finding elongational flow events (characterized by $\Q>0.9$) increases.
This behavior reflects the action of the Marangoni forces, whose amplitude is proportional to the elasticity number.
Indeed, larger elasticity numbers (stronger surfactants) lead to stronger surface tension reductions and thus larger surface tension gradients (i.e. larger Marangoni forces).
Marangoni forces, which are tangential to the interface and are directed as the surface tension gradients, generate a flow similar to an elongational flow (see figure~\ref{fig: sketch}) and thus for increasingly larger elasticity numbers, the probability of finding elongational flow events is larger.
For $\beta_s=4.00$, an additional effect enters the picture and the trend is lost.
This effect is linked to the shutdown of Marangoni forces: for surfactant concentrations above the shutdown concentration, $\psi_s$, surface tension keeps constant and Marangoni forces vanish.
For the strongest surfactant (i.e. $\beta_s=4.00$), the shutdown concentration is extremely low, $\psi_s\simeq0.12$, thus most of the interface is characterized by a lower, but constant, surface tension, $\sigma/\sigma_0=0.5$.
As Marangoni forces vanish, the flow inside the droplets is less likely to be extensional and thus the trend is lost.
This effect is also partially visible for $\beta_s=2.0$ (inset of figure~\ref{fig: Q_drop}{a}): the increase in the probability of having elongational flow inside the droplets is still higher than the cases at $\beta_s=1.0$, but the difference between the two cases is rather small.
The shift of the PDFs towards rotational and pure shear flow events ($\Q \le 0$) for increasing elasticity numbers is determined also by two additional factors: the smaller droplet size and the average surface tension reduction.
These two factors promote the presence of rotational flow (smaller droplet size) and of pure shear flow regions (average surface tension reduction) \cite{RostiGJDB_2019}.

At $We=3.00$ (figure~\ref{fig: Q_drop}b), the PDFs of the flow topology parameter exhibit some remarkable changes with respect to $We=1.50$: for all the cases, the probability of observing rotational and shear flow events increases, while that of observing elongational flow events decreases.
These modifications can be linked to the two above mentioned factors (smaller droplet size and average surface tension reduction), which lead to a larger probability of finding rotational and pure shear flow regions. 
In particular, the smaller droplet size introduces a strong confinement effect: the flow inside the droplets is more confined by the interface and thus the internal flowing is suppressed.
As the internal flow becomes more confined, the droplets are less subjected to the mean shear imposed by the external flow and rotational flow events are promoted \cite{RostiGJDB_2019}.
Similarly, the lower surface tension of these cases reducing the damping capability of the interface and the internal flow field becomes more similar to that of the carrier phase (dominated by pure shear flow events).
These two effects are in competition with the Marangoni forces, which, by opposite, favor an elongational type of flow.
The outcome of this competition can be appreciated by comparing the PDFs obtained from the different elasticity numbers.
While for $\beta_s=0.50$ and $\beta_s=1.00$ the probability of finding elongational flow events increases suggesting the dominance of the Marangoni forces, for $\beta_s=2.00$ and $\beta_s=4.00$ the trend is lost and a larger probability of finding rotational and pure shear flow events is observed.
This latter observation, which can be only partially linked to the shutdown of the Marangoni forces \cite{Soligo2019_jfm}, indicates that for these two latter cases the effects induced by the smaller droplet size and surface tension reduction are predominant.
As a consequence, the PDFs shift towards rotational and pure shear flow events.

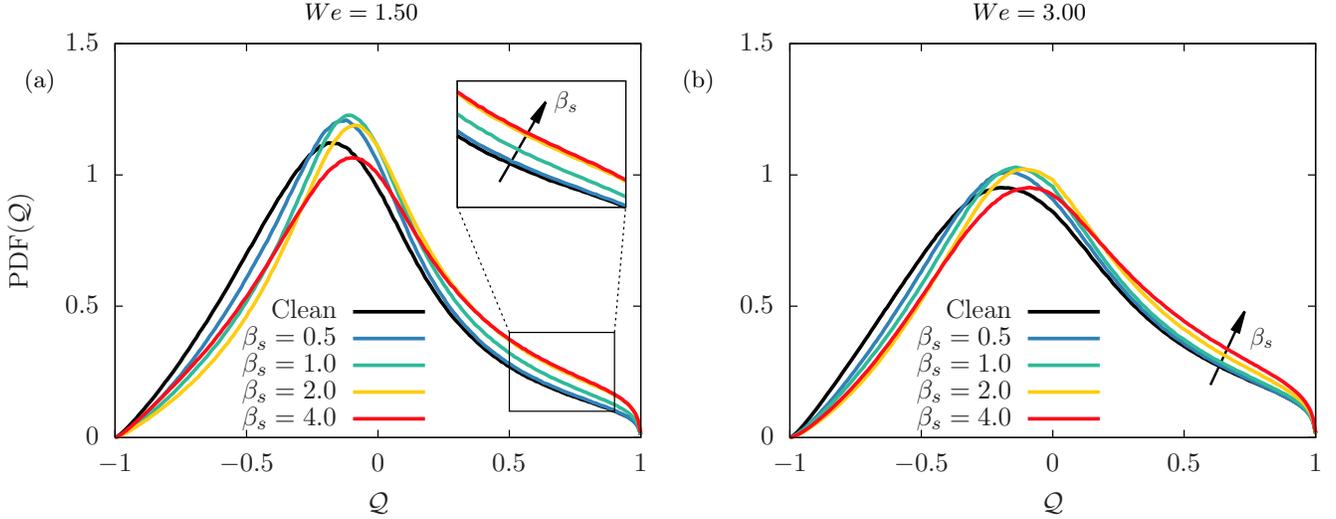
\begin{figure}
\center
\setlength{\unitlength}{0.0025\columnwidth}
\begin{picture}(400,150)
\put(-20,-25){\includegraphics[width=0.65\columnwidth, keepaspectratio]{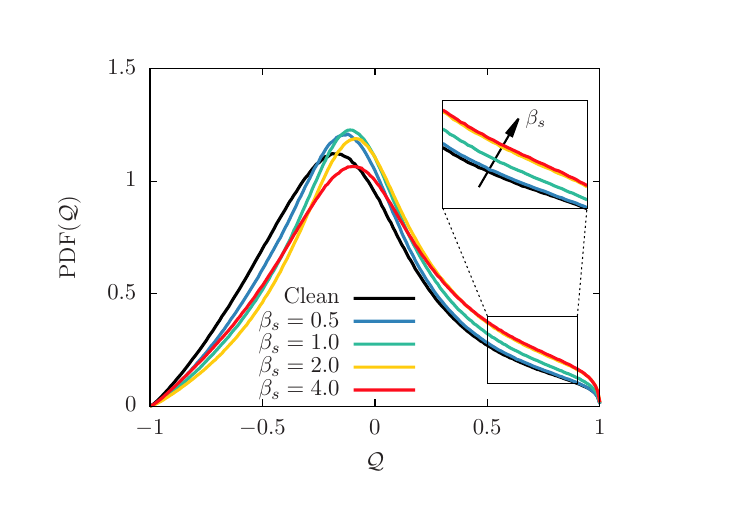}}
\put(180,-25){\includegraphics[width=0.65\columnwidth, keepaspectratio]{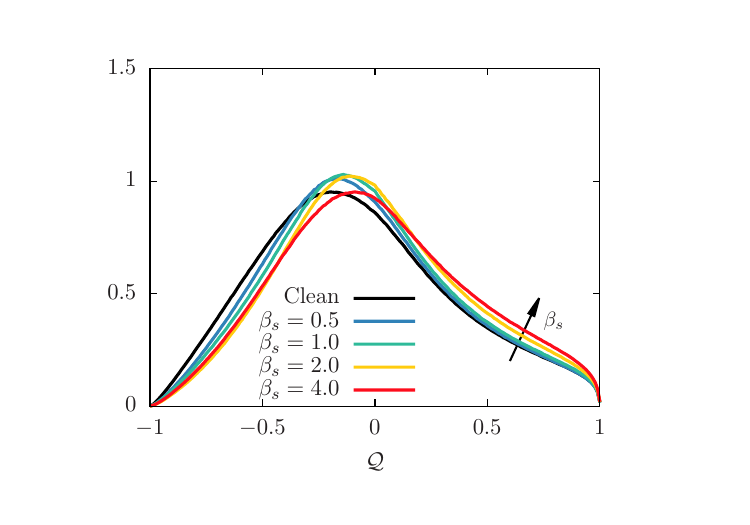}}
\put(5,120){(a)}
\put(200,120){(b)}
\put(88,140){$We=1.50$}
\put(286,140){$We=3.00$}
\end{picture}
\caption{Probability density function (PDF) of the flow topology parameter $\Q$ at the interface.
Panel (a) refers to $We=1.50$, while panel (b) to $We=3.00$.
The different cases are reported using different colors: clean (black), $\beta_s=0.50$ (blue), $\beta_s=1.00$ (green), $\beta_s=2.00$ (yellow) and $\beta_s=4.00$ (red).
For both Weber numbers, the probability of observing elongational flow events increases for increasing elasticity numbers  and, by opposite, the probability of finding rotational flow events decreases.}
\label{fig: Q_interf}
\end{figure}

Finally, we consider the PDF of the flow topology parameter at the interface, figure~\ref{fig: Q_interf}.
Panel (a) refers to $We=1.50$ while panel (b) refers to $We=3.00$. The different cases are plotted with different colors: clean (black), $\beta_s=0.50$ (blue), $\beta_s=1.00$ (green), $\beta_s=2.00$ (yellow) and $\beta_s=4.00$ (red).
The computation of the statistics only at the interface (and thus excluding the internal part of the droplets) enables us to sort out the effects induced by the smaller droplet size and by the surface tension reduction.

We start by considering the behavior of the PDFs in the region $0<\Q<+1$.
For both Weber numbers, increasing the elasticity number (i.e. increasing the surfactant strength), we observe a larger probability to find elongational flow events.
This observation reflects the action of the Marangoni forces (whose magnitude increases with the elasticity number), that generate an elongational type of flow.
In addition, the effect of the shutdown of Marangoni forces is also visible: for $We=1.50$ it involves a considerable part of the surfactant at the interface and thus the difference between the cases $\beta_s=2.00$ and $\beta_s=4.00$ is very small (see inset of figure~\ref{fig: Q_interf}a).
Conversely, for $We=3.00$, the results are less influenced by the shutdown of the Marangoni forces (the larger interfacial area leads to a lower average surfactant concentration \cite{Soligo2019_jfm}) and  the trend is kept.

Turning to the behavior of the PDFs in the region $-1<\Q<0$, for all multiphase cases (expect for  $\beta_s=4.00$), we observe a decreasing probability of observing rotational-like flow events for increasing elasticity numbers (surfactant strengths).
This indicates that the surface tension forces (especially the capillary ones) generate a rotational type of flow at the interface.
Indeed, the amplitude of these forces is larger when surfactant-free systems or weaker surfactants are considered and vorticity tangential to the interface is generated (see figure~\ref{fig: vort}).
Focusing on the cases at $\beta_s=4.00$ (especially at $We=1.50$), the trend is partially lost since the shutdown of the Marangoni forces strongly limits the occurrence of elongation flow events.
Thus, the resulting PDF is somehow more similar to that obtained from a surfactant-free case (absence of Marangoni forces).
This similarity can be appreciated comparing the case $We=1.50$ and $\beta_s=4.00$ (figure~\ref{fig: Q_interf}a, red) and the surfactant-free case at $We=3.00$ (figure~\ref{fig: Q_interf}a, black), which indeed are characterized by a similar equivalent Weber number, $We \simeq 2.70$ for the first case and $We=3.00$ for the second.
In addition, for the first case, the shutdown of the Marangoni forces involves a wider portion of the interface (i.e. capillary forces are predominant) explaining the similarity between the two PDFs.

A detailed discussion on the effect of the Reynolds number on the PDF of the flow topology parameter is reported in appendix~\ref{appa}, where the same statistic has been analyzed at a lower Reynolds number.

\subsection{Effect of grid resolution on the results}
\label{igrid}

In this last section, we would like to address the effect of the grid resolution on the local flow indicators obtained. 
In the framework of phase-field approaches, the thickness of the interfacial transition layer is set via the Cahn number; it is, however, good practice to select the Cahn number according to the grid resolution.
Thus, here we investigate the effect of grid resolution and interface thickness (i.e. Cahn number) on the alignment between vorticity and interface, and on the flow topology parameter sampled at the interface.
%
To this end, we consider the case at $We=3.00$ and $\beta_s=4.00$, and we compare the results obtained from two different grid resolutions (and, thus, two different Cahn numbers):  $N_x \times N_y \times N_z=1024\times512\times513$, (standard grid used for all cases, $Ch=0.025$) and $N_x \times N_y \times N_z=2048 \times 1024 \times 1025$ (finer grid, refined twice in all directions, $Ch=0.0125$).

\begin{figure}
\center
\setlength{\unitlength}{0.0025\columnwidth}
\begin{picture}(400,150)
\put(-20,-25){\includegraphics[width=0.65\columnwidth, keepaspectratio]{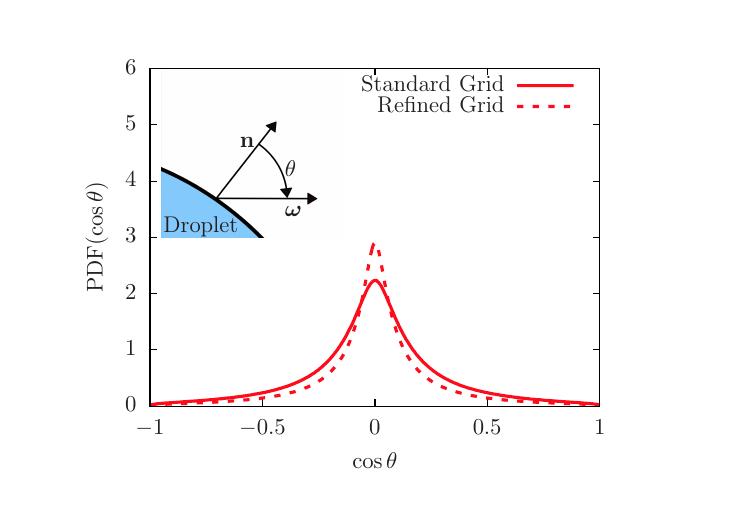}}
\put(180,-25){\includegraphics[width=0.65\columnwidth, keepaspectratio]{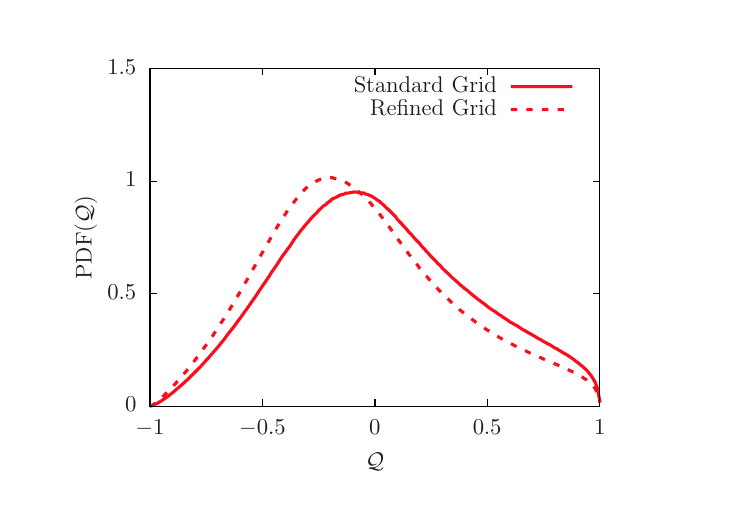}}
\put(5,120){(a)}
\put(200,120){(b)}
\end{picture}
\caption{Probability density function (PDF) of  $\cos\theta$ computed at the interface (panel a) and probability density function (PDF) of the flow topology parameter $\Q$ sampled at the interface (panel b).
Both panels refer to the case $We=3.00$ and $\beta_s=4.00$.
Results obtained using the standard grid ($N_x \times N_y \times N_z=1024\times512\times513$) and $Ch=0.025$ are reported with solid lines, while those obtained from the refined grid simulation ($N_x \times N_y \times N_z=2048 \times 1024 \times 1025$) and $Ch=0.0125$ are reported with dashed lines.}
\label{fig: grid}
\end{figure}

In figure~\ref{fig: grid}, we show the probability density function (PDF) of  $\cos\theta$ (panel a) and of the flow topology parameter $\Q$ at the interface (panel b) for the two different grid resolutions considered.
The results obtained from the standard grid simulation are reported with solid lines, while those obtained from the refined grid simulation are reported with dashed lines.

Considering the PDF of $\cos \theta$ (figure~\ref{fig: grid}a), we observe that increasing the grid resolution (and consequently decreasing the Cahn number), the probability of having the vorticity unit vector perpendicular to the interface normal slightly increases \cite{shao2018direct}.
This modification of the PDF can be traced back mainly to two effects: as the grid resolution is increased, smaller droplets can be better captured and the interfacial layer is thinner.
Smaller droplets are characterized by a very low deformability (i.e. a smaller Weber number, computed using the droplet diameter as a reference length scale) and, thus, produce stronger modifications of the local flow field. 
Indeed, as the droplet becomes less deformable, the probability of finding vorticity perpendicular to the interface normal increases.
The second effect (thinner interfacial layer) leads to a local increase in surface tension forces: while the integral of surface tension forces across the interfacial layer is the same for both cases, when the Cahn number is halved (i.e. the interface thickness is halved), surface tension forces are smeared out on a  thinner interfacial layer and are thus locally higher. 

Similar effects can be appreciated as well for the PDF of the flow topology parameter sampled at the interface (figure~\ref{fig: grid}b): smaller droplets suppress internal flowing (i.e. reduce elongational-like events), while higher surface tension forces increase the stiffness of the interface (i.e. increase pure shear flow events, which can be appreciated by comparing panels a-b of figure~\ref{fig: Q_interf}).
Indeed, as the thickness of the interfacial layer is reduced (lower Cahn number, refined grid), the peak of the PDF shifts towards smaller values of $\mathcal{Q}$.

The weak dependence on the grid resolution of the results presented above can be justified as follows: as the grid resolution is increased, the dynamics of smaller structures (e.g. smaller droplets and ligaments) can be captured.
Whether or not these smaller structures affect the results depends on the statistic considered: they have a negligible effect on macroscopic flow indicators, while a slightly more pronounced effect is observed on local flow indicators and on droplet statistics \cite{vincent2019phase,shao2018direct,Soligo2019_jfm}. 
Thus, for the interface-vorticity alignment and the flow topology parameter, which are local quantities computed at the interface (whose shape and extension depend on the dispersed phase morphology), a weak dependence of the results on the grid resolution can be expected and has been also observed in previous studies \cite{shao2018direct}.

\section{Conclusions}
\label{sec: conclusions}

In this work, we investigated the modifications on the flow field induced by a swarm of surfactant-laden droplets.
The dynamics of the multiphase flow is described via direct numerical simulations of the Navier-Stokes equation coupled with a two-order-parameter phase-field method.
The first order parameter, the phase field, describes the local concentration of the carrier and dispersed phase and the second order parameter defines the surfactant concentration.
The feedback of the surfactant-laden interface onto the flow field is accounted for via a force-coupling term in the Navier-Stokes equations.
The effects of both Weber number (ratio of inertial forces over surface tension forces) and elasticity number (surfactant strength) on both global and local flow parameters have been investigated.
We found that the macroscopic flow is almost unaffected by the presence of the surfactant-laden droplets since the total volume fraction of the dispersed phase is rather low ($\Phi \simeq 5.4\%$), as also observed at lower Reynolds number by Scarbolo {\em et al.} 2016 \cite{Scarbolo2016} and Roccon {\em et al.} 2017 \cite{Roccon2017}.
However, we could observe a mild suppression of the turbulence fluctuations (streamwise and wall-normal) in the channel core, where the majority of the droplets are located.
By opposite, remarkable modifications of the local flow field are observed.
The vorticity field at the interface is influenced by the presence of surface tension forces and a larger probability of finding vorticity tangential at the interface is obtained when lower Weber and elasticity numbers are considered (i.e. when larger surface tension values are considered).
Large modifications are observed also for the flow topology parameter $\Q$, which we compute separately in the carrier phase, dispersed phase and at the interface.
In the carrier phase, the flow is dominated by the mean shear arising from the two solid walls bounding the flow and, thus, mainly pure shear flow events ($\Q=0$) are observed, as in a canonical single-phase flow.
Differently, the dispersed phase is characterized by a combination of rotational ($\Q=-1$), pure shear ($\Q=0$) and elongational ($\Q=+1$) flow regions.
Elongational flow regions are predominant for $We=1.50$, while the probability of observing rotational and pure shear flow regions increases for the cases at $We=3.00$. 
The different flow topology observed in the dispersed phase is the result of the competition between different factors: the Marangoni forces, which promote an elongational flow, and the smaller droplet size and surface tension reduction, which promote the presence of rotational and pure shear flow regions.
At the interface, the action of the Marangoni forces on the flow topology parameter can be clearly observed: for both Weber numbers, increasing the elasticity number (i.e. increasing the magnitude of the Marangoni forces) increases the  probability of finding elongational flow regions, while the probability of observing rotational flow regions decreases.

\begin{acknowledgments}
We acknowledge PRACE (Project ID: 2018194645, 30M core hours on Marconi-KNL) and ISCRA (Project ID: HP10BOR3UN, 10M core hours on Marconi-KNL) for the generous allowance of computational resources. GS gratefully acknowledges funding from the PRIN project \textit{Advanced Computations and Experiments in Turbulent Multiphase Flow} (project code 2017RSH3JY) and from the MSCA-ITN-EID project \textit{COMETE} (project code 813948).
\end{acknowledgments}

\appendix
\section{Effect of the Reynolds number}
\label{appa}

Three additional simulations were performed to investigate the effect of the Reynolds number on the quantities of interest.
In particular, two droplet-laden cases (clean droplets at different Weber numbers, $We=1.50$ and $We=3.00$) and one reference single-phase case at shear Reynolds $Re_\tau=150$ have been performed. 
The very same numerical setup has been adopted: the boundary and initial conditions, the grid resolution ($N_x \times N_y \times N_z = 1024 \times 512 \times 513$) and the phase-field parameters ($Ch=0.025$ and $Pe_\phi=150$) are unchanged with respect to the cases presented in table~\ref{overview}.
At first, the effect of the Reynolds number on the macroscopic indicators will be presented, then the local flow indicators will be investigated.

\subsection{Macroscopic flow modifications}
\begin{figure}
\center
\setlength{\unitlength}{0.0025\columnwidth}
\begin{picture}(400,150)
\put(80,-25){\includegraphics[width=0.65\columnwidth, keepaspectratio]{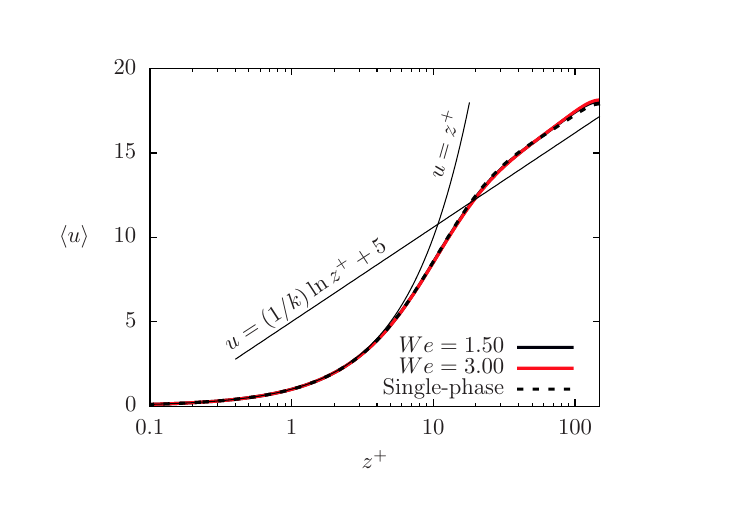}}
\end{picture}
\caption{Profiles of the mean streamwise velocity, $\langle u \rangle$, for the simulations performed at $Re_\tau=150$: $We=1.50$ (black solid line) and $We=3.00$ (red solid line); the $z^+$ axis is reported in log scale (base 10).
The single-phase velocity profile is reported with a black dashed line.
The classical law of the wall, $u=z^+$ and $u=(1/k) \ln z^+ +5$ (being $k=0.41$ the von K{\'a}rm{\'a}n constant), is also reported as a reference.}
\label{fig: u_mean_re}
\end{figure}

Similarly to what observed at the higher shear Reynolds number, $Re_\tau=300$, the relatively low volume fraction of the dispersed phase (about 5\%) limits its effect on the macroscopic flow.
In particular, a slight increase in the mean streamwise velocity profile, figure~\ref{fig: u_mean_re}, is observed also at the lower shear Reynolds number, $Re_\tau=150$, in the core region of the channel. 
The mean streamwise velocity profile for both Weber numbers are superposed: for the considered volume fraction, the Weber number has no influence on the mean velocity.

\begin{figure}
\center
\setlength{\unitlength}{0.0025\columnwidth}
\begin{picture}(400,150)
\put(-20,-25){\includegraphics[width=0.65\columnwidth, keepaspectratio]{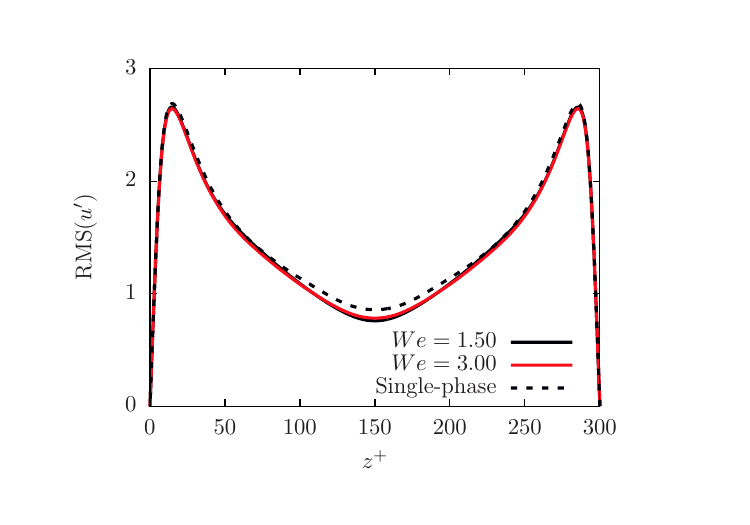}}
\put(180,-25){\includegraphics[width=0.65\columnwidth, keepaspectratio]{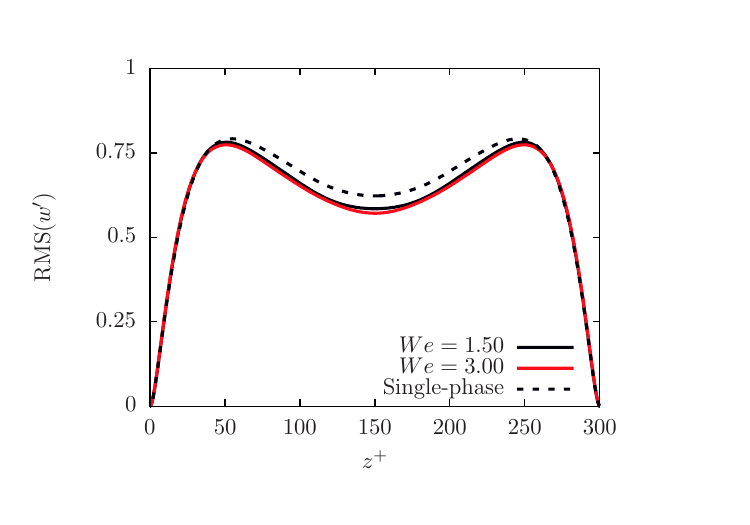}}
\put(5,120){(a)}
\put(200,120){(b)}
\end{picture}
\caption{Root mean square (RMS) of the velocity fluctuations for the simulations performed at $Re_\tau=150$: streamwise component (panel a) and wall-normal component (panel b). 
The different cases are shown using different line styles: $We=1.50$ (black solid lines) and $We=3.00$ (red solid lines).
Black dashed lines identify the single-phase flow results.}
\label{fig: uw_rms_re}
\end{figure}

A decrease in the amplitude of velocity fluctuations in channel core (streamwise and wall-normal components, figure~\ref{fig: uw_rms_re}, respectively panels a-b) is observed for both the shear Reynolds number values considered ($Re_\tau=150$ in figure~\ref{fig: uw_rms_re} and $Re_\tau=300$ in figure~\ref{fig: uw_rms}); the magnitude of the reduction is similar for both $Re_\tau$ values.
Indeed, the dispersed phase introduces a mild turbulence modulation in the core of the channel; a stronger effect is observed for the wall-normal component.

\subsection{Local flow modifications}
\begin{figure}
\center
\setlength{\unitlength}{0.0025\columnwidth}
\begin{picture}(400,150)
\put(88,140){$We=1.50$}
\put(286,140){$We=3.00$}
\put(-20,-25){\includegraphics[width=0.65\columnwidth, keepaspectratio]{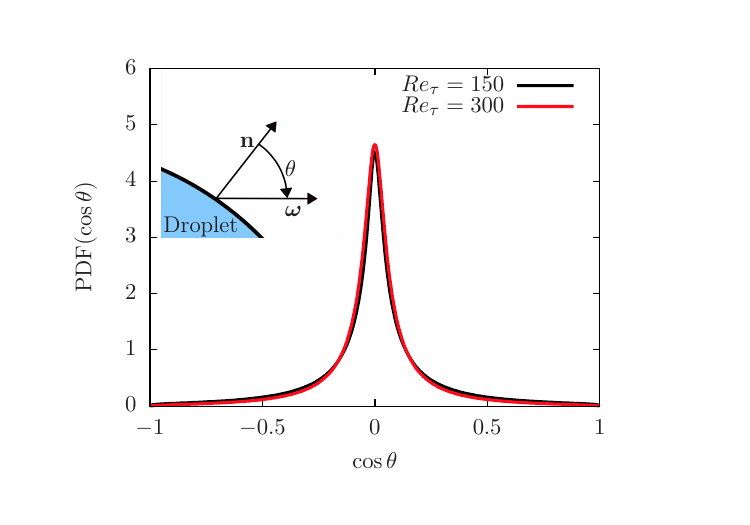}}
\put(180,-25){\includegraphics[width=0.65\columnwidth, keepaspectratio]{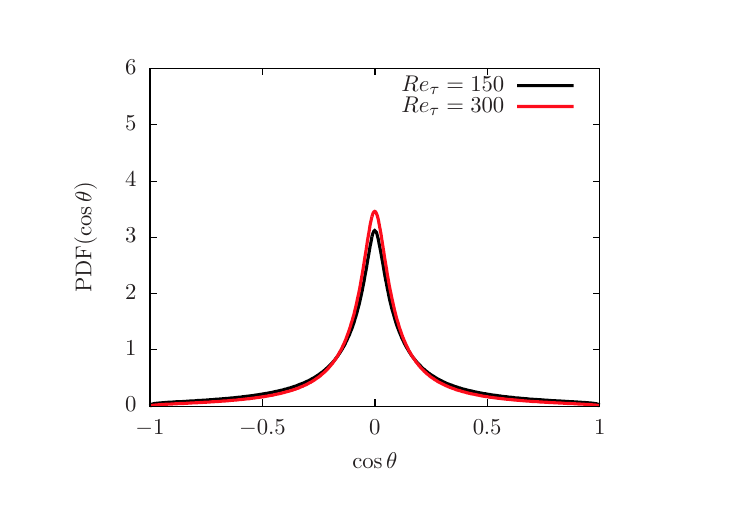}}
\put(5,120){(a)}
\put(200,120){(b)}
\end{picture}
\caption{Probability density function (PDF) of  $\cos\theta$ computed at the interface different Reynolds numbers: $Re_\tau=150$ (black solid lines) and $Re_\tau=300$ (red solid lines). Panel (a) reports the data at $We=1.50$, while panel (b) those at $We=3.00$.}
\label{fig: vort_re}
\end{figure}
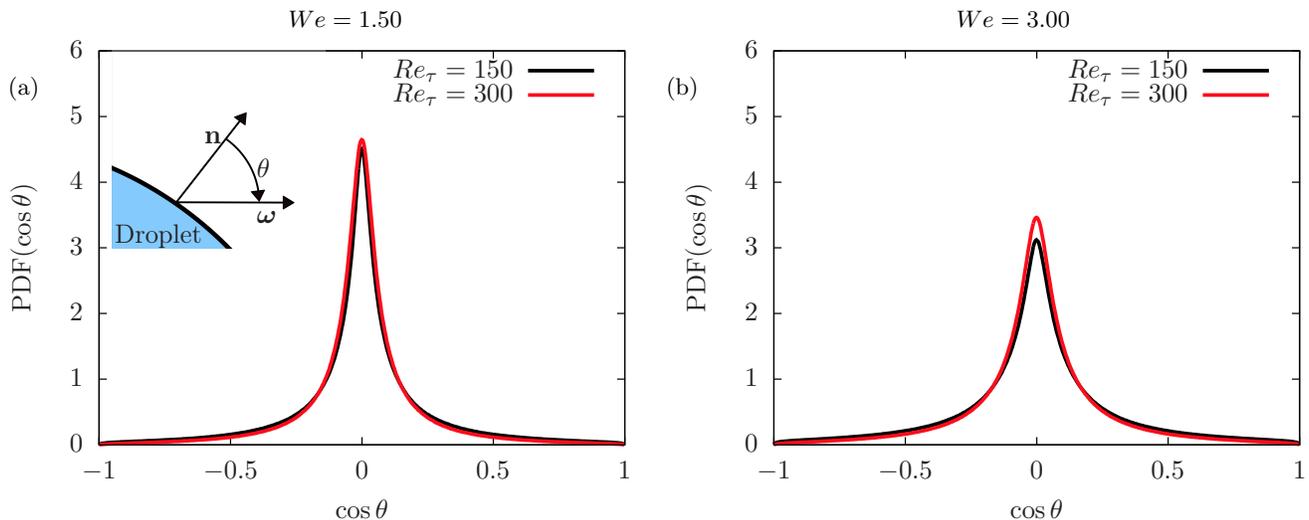

The alignment between the vorticity at the interface and the interface normal is almost unchanged for different shear Reynolds numbers: figure~\ref{fig: vort_re} shows the PDF of $\cos\theta$, being $\theta$ the angle between the vorticity $\boldsymbol{\omega}$ and the interface normal $\mathbf{n}$.
The PDFs for the two shear Reynolds numbers are almost superposed at the lower Weber number, $We=1.50$, while at $We=3.00$ a stronger preferential alignment is observed for the higher shear Reynolds number, $Re_\tau=300$.
Indeed, at higher shear Reynolds number the vorticity is more likely to be orthogonal to the interface normal. 
This difference can be traced back to the different dispersed phase morphology observed when the shear Reynolds number is increased: an increase in the shear Reynolds number leads to a reduction in the Hinze diameter \cite{Hinze1955} through an increase in the dissipation rate, thus lowering the average droplet size and moving the droplet size distribution towards smaller diameters. 
This effect is clear for the cases at $We=3.00$ (lower surface tension), while for the cases at $We=1.50$ the higher surface tension keeps the larger droplets stable despite the increase in the dissipation rate. 
This way, almost no shear Reynolds number effect can be observed for the low-Weber number case, while a minor effect is observed for the high-Weber number case.
The smaller droplets that are generated when the shear Reynolds number is increased (cases at $We=3.00$) have a higher curvature and, thus, are less deformable.
 Indeed, a stronger preferential alignment at $\theta=90^\circ$ is observed when the shear Reynolds number is increased.
Thus, the shear Reynolds number has an indirect effect on the alignment of vorticity at the interface: by increasing the dissipation rate (hence decreasing the value of Hinze diameter), it changes the morphology of the dispersed phase, leading to the formation of smaller and less deformable droplets.
In turn, these smaller droplets favor the production of vorticity perpendicular to the interface normal (i.e. $\cos \theta = 0$).

\begin{figure}
\center
\setlength{\unitlength}{0.0025\columnwidth}
\begin{picture}(400,150)
\put(-20,-25){\includegraphics[width=0.65\columnwidth, keepaspectratio]{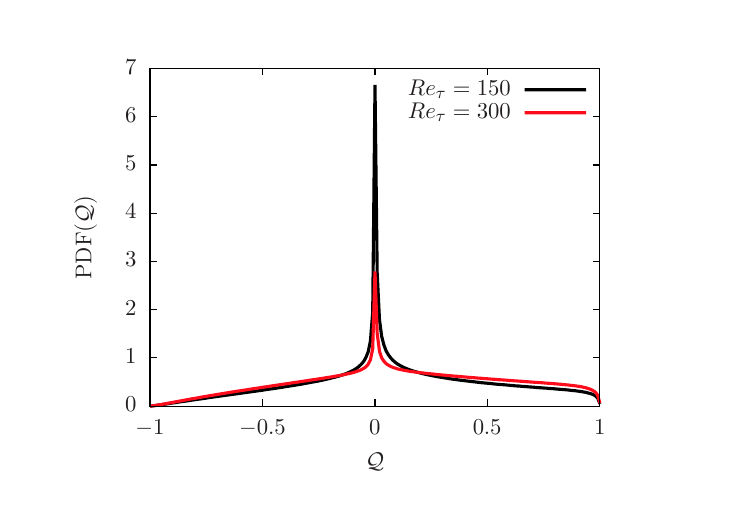}}
\put(180,-25){\includegraphics[width=0.65\columnwidth, keepaspectratio]{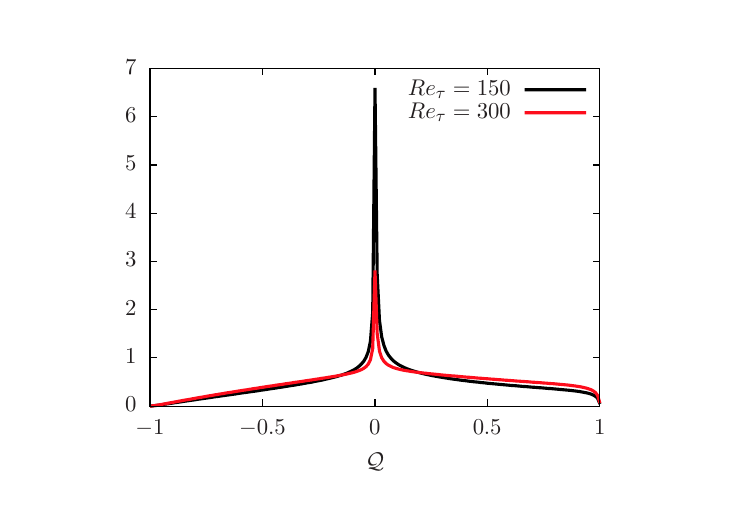}}
\put(5,120){(a)}
\put(200,120){(b)}
\put(88,140){$We=1.50$}
\put(286,140){$We=3.00$}
\end{picture}
\caption{Probability density function (PDF) of the flow topology parameter $\Q$ in the carrier phase for different Reynolds numbers: $Re_\tau=150$ (black solid lines) and $Re_\tau=300$ (red solid lines). Panel (a) reports the data at $We=1.50$, while panel (b) those at $We=3.00$.}
\label{fig: Q_carrier_re}
\end{figure}
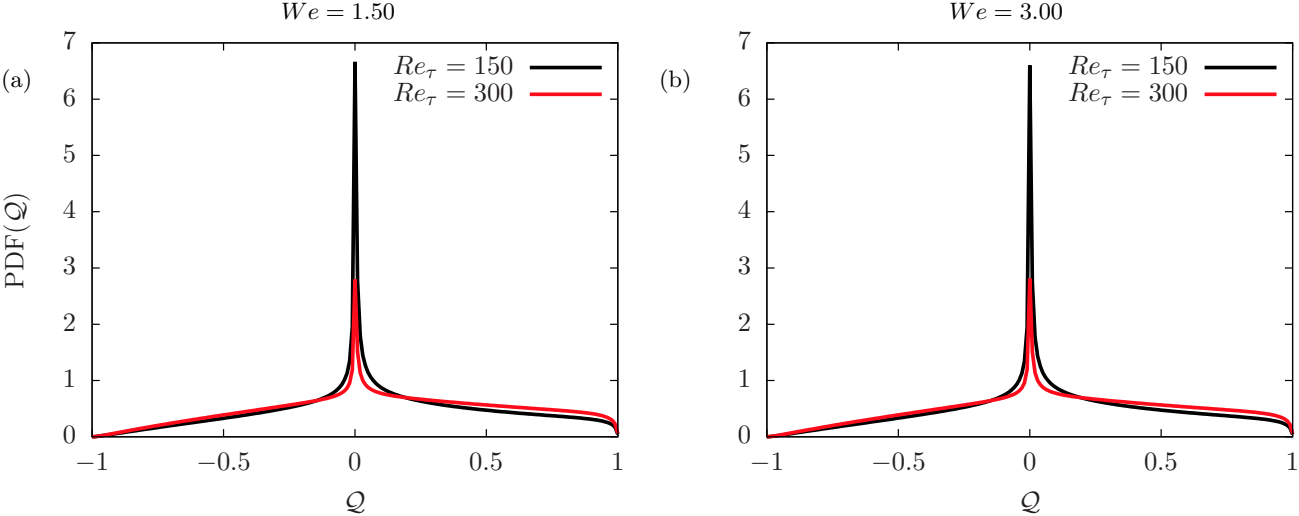

A relevant effect of the shear Reynolds number is observed when the flow topology parameter is considered (figure~\ref{fig: Q_carrier_re}). 
In particular, the carrier phase is characterized by a strong reduction in pure shear and an increase in elongational type of flow. 
The increase in the pure shear contribution observed for the lower shear Reynolds number is linked to the higher fraction of the channel occupied by the viscous sublayer, with respect to the case at $Re_\tau=300$.
The thickness of the viscous sublayer keeps the same when the shear Reynolds number is changed (about 10 wall units), but the total channel height is different: 300 wall units at $Re_\tau=150$ and 600 wall units at $Re_\tau=300$.
Therefore, the presence of a dispersed phase has almost no effect on the flow topology parameter in the carrier phase, which is instead dominated by the imposed flow, and the differences observed are to be attributed to the different extension of the viscous sublayer with respect to the total size of the channel.

\begin{figure}
\center
\setlength{\unitlength}{0.0025\columnwidth}
\begin{picture}(400,150)
\put(-20,-25){\includegraphics[width=0.65\columnwidth, keepaspectratio]{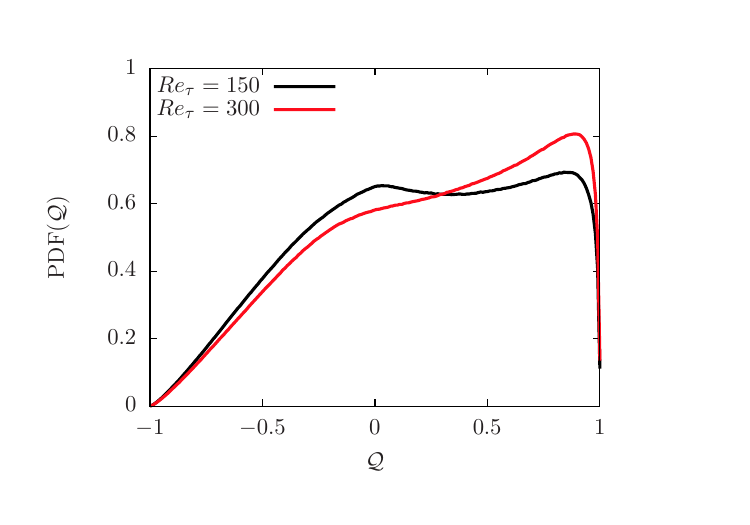}}
\put(180,-25){\includegraphics[width=0.65\columnwidth, keepaspectratio]{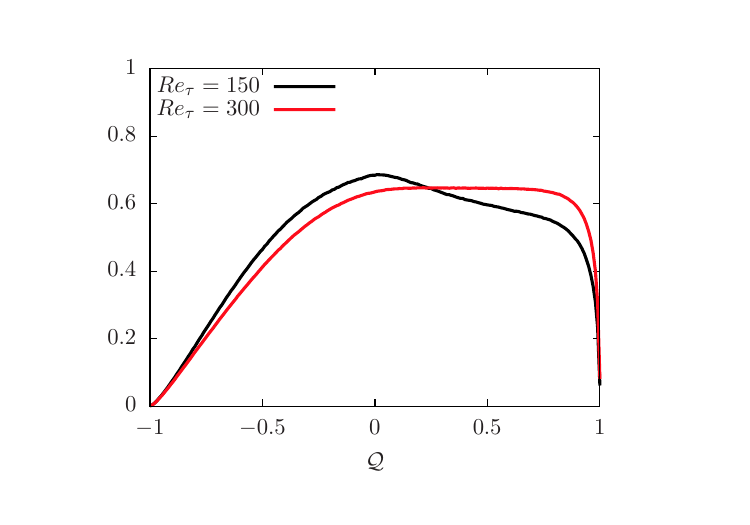}}
\put(5,120){(a)}
\put(200,120){(b)}
\put(88,140){$We=1.50$}
\put(286,140){$We=3.00$}
\end{picture}
\caption{Probability density function (PDF) of the flow topology parameter $\Q$ in the disperse phase for different Reynolds numbers: $Re_\tau=150$ (black solid lines) and $Re_\tau=300$ (red solid lines). Panel (a) reports the data at $We=1.50$, while panel (b) those at $We=3.00$.}
\label{fig: Q_drops_re}
\end{figure}
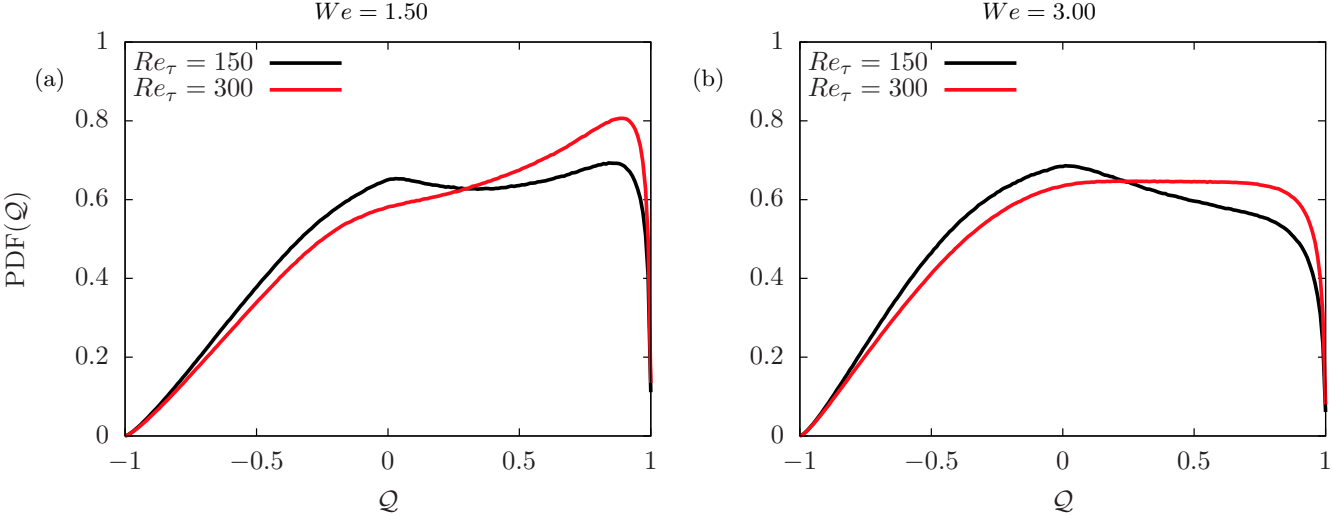

The dispersed phase (figure~\ref{fig: Q_drops_re}), however, is mainly located in the inner layer, thus it is mostly unaffected by the viscous sublayer (as can be confirmed by the absence of a strong peak of pure shear). 
Again, the main differences observed are to be linked to the different imposed flow: as the shear Reynolds number is increased, the PDF of the flow topology parameter (in the carrier phase or for a single-phase flow case) shows a reduction in pure shear type of flow and an increase in elongational type of flow. 
This change is then reflected in the PDF of the flow topology parameter, figure~\ref{fig: Q_drops_re}, which exhibits a higher probability of finding elongational type of flow inside the droplets.
The reduction in the probability of elongational type of flows inside the droplets for the lower shear Reynolds number is also linked to a stronger internal flow suppression: the droplets are roughly of similar size (when comparing data at the same Weber number), but viscous effects are stronger (i.e. lower shear Reynolds number).

\begin{figure}
\center
\setlength{\unitlength}{0.0025\columnwidth}
\begin{picture}(400,150)
\put(-20,-25){\includegraphics[width=0.65\columnwidth, keepaspectratio]{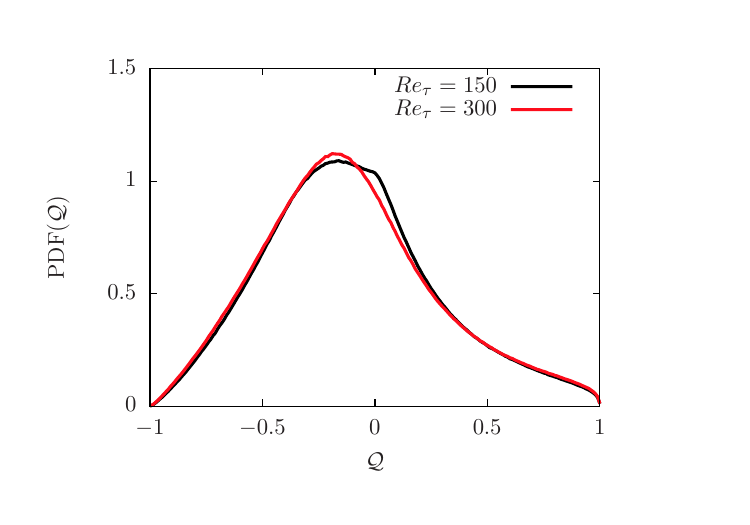}}
\put(180,-25){\includegraphics[width=0.65\columnwidth, keepaspectratio]{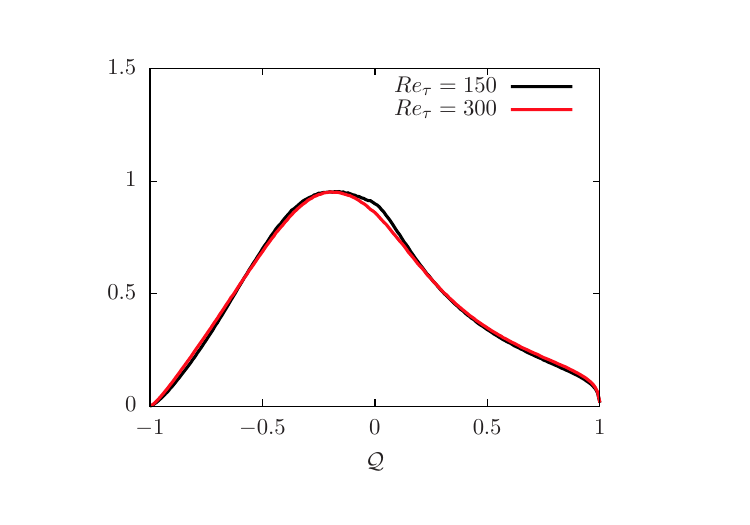}}
\put(5,120){(a)}
\put(200,120){(b)}
\put(88,140){$We=1.50$}
\put(286,140){$We=3.00$}
\end{picture}
\caption{Probability density function (PDF) of the flow topology parameter $\Q$ at the interface for different Reynolds numbers: $Re_\tau=150$ (black solid lines) and $Re_\tau=300$ (red solid lines). Panel (a) reports the data at $We=1.50$, while panel (b) those at $We=3.00$.}
\label{fig: Q_inter_re}
\end{figure}
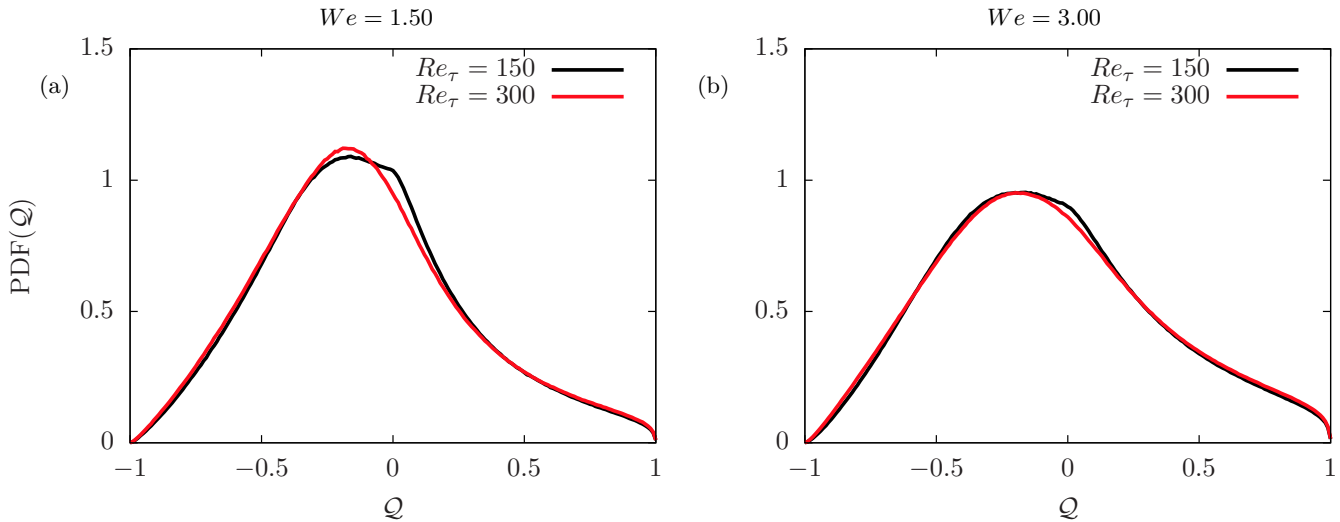

Similar considerations to those done for effect of the shear Reynolds number on the flow topology parameter in the dispersed phase still apply at the interface, figure~\ref{fig: Q_inter_re}: an increase in the probability of pure shear and a decrease in the probability of elongational flow is observed as the shear Reynolds number is decreased. 
Indeed, as the shear Reynolds number is increased the probability of elongational and rotational flow increases, while that of pure shear decreases. 
Here, however, the confinement effect is filtered out (the PDF is computed only at the interface), hence the effect of the shear Reynolds number is much less marked.

\end{document}